\documentclass[%
 reprint,
superscriptaddress,
 amsmath,amssymb,
 aps,
]{revtex4-1}
\usepackage{amsmath}
\usepackage{graphicx,epstopdf}% Include figure files
\usepackage{dcolumn}% Align table columns on decimal point
\usepackage{bm}% bold math
\usepackage{xcolor}
\usepackage{float}
\usepackage{enumerate}
\usepackage{natbib}
\usepackage{multirow}
\begin{document}

%\preprint{APS/123-QED}

\title{Thermal conductivity of amorphous and crystalline GeTe thin film at high temperature: Experimental and Theoretical study}% Force line breaks with \\
%\thanks{A footnote to the article title}%

\author{Kanka Ghosh}
\email{Corresponding author: kanka.ghosh@u-bordeaux.fr}
\affiliation{University of Bordeaux, CNRS, Arts et Metiers Institute of Technology, Bordeaux INP, INRAE, I2M Bordeaux, F-33400 Talence, France}

 %\altaffiliation[Also at ]{Physics Department, XYZ University.}%Lines break automatically or can be forced with \\
\author{Andrzej Kusiak}%
\affiliation{University of Bordeaux, CNRS, Arts et Metiers Institute of Technology, Bordeaux INP, INRAE, I2M Bordeaux, F-33400 Talence, France}
 %\email{Second.Author@institution.edu}
 \author{Pierre No\'e}
 \affiliation{CEA-LETI, MINATEC, Grenoble, France}
 \author{Marie-Claire Cyrille}
 \affiliation{CEA-LETI, MINATEC, Grenoble, France}
 \author{Jean-Luc Battaglia}%
 %\email{jean-luc.battaglia@u-bordeaux.fr}
 \affiliation{University of Bordeaux, CNRS, Arts et Metiers Institute of Technology, Bordeaux INP, INRAE, I2M Bordeaux, F-33400 Talence, France}
 %\altaffiliation{jean-luc.battaglia@u-bordeaux.fr}
 %\email{third.Author@institution.edu}
%

%\collaboration{CLEO Collaboration}%\noaffiliation

%\date{}% It is always \today, today,
             %  but any date may be explicitly specified

\begin{abstract}
Thermal transport properties bear a pivotal role in influencing the performance of phase change memory (PCM) devices, in which, the PCM operation involves fast and reversible phase change between amorphous and crystalline phases. In this article, we present a systematic experimental and theoretical study on the thermal conductivity of GeTe at high temperatures involving fast-change from amorphous to crystalline phase upon heating. Modulated photothermal radiometry (MPTR) is used to experimentally determine thermal conductivity of GeTe at high temperatures in both amorphous and crystalline phases. Thermal boundary resistances are accurately taken into account for experimental consideration. To develop a concrete understanding of the underlying physical mechanism, rigorous and in-depth theoretical exercises are carried out. For this, first-principles density functional methods and linearized Boltzmann transport equations (LBTE) are employed using both direct and relaxation time based approach (RTA) and compared with that of the phenomenological Slack model. The amorphous phase experimental data has been described using the minimal thermal conductivity model with sufficient precision. The theoretical estimation involving direct solution and RTA method are found to retrieve well the trend of the experimental thermal conductivity for crystalline GeTe at high temperatures despite being slightly overestimated and underestimated respectively compared to the experimental data. A rough estimate of vacancy contribution has been found to modify the direct solution in such a way that it agrees excellently with the experiment. Umklapp scattering has been determined as the significant phonon phonon scattering process. Umklapp scattering parameter has been identified for GeTe for the whole temperature range which can uniquely determine and compare umklapp scattering processes for different materials.
\end{abstract}

%\keywords{Suggested keywords}%Use showkeys class option if keyword
                              %display desired
\maketitle

%\tableofcontents
\section{Introduction}

Chalcogenide alloys have been evolved as excellent candidates for
the purpose of electronic nonvolatile memory storage (phase change
memories-PCM)\cite{Kolobov,Wuttig,Wang2008,Lu2013,Campi2015,Campi2017,Andrzej}. This application
involves a fast and reversible alteration between amorphous and crystalline
phase on heating. PCM cells consist of a nanoscale volume of a phase
change (PC) material, normally a Tellurium (Te) based alloy, which
undergoes a reversible change between amorphous and crystalline states,
possessing a contrasting electrical resistivity and thus enabling the PCMs
to be used for binary data storage \cite{Kolobov,Lu2013}. In a PCM device,
crystallization by heating the amorphous PC alloy above its crystallization
temperature with electric current pulses is called SET operation while
amorphization of the crystalline region by melting and quenching using
higher and shorter electric current pulses is called RESET operation
\cite{Andrzej}. Germanium telluride (GeTe) is one of the promising candidates within the phase change materials due to its notably high contrast in electrical resistance as well as a
stable amorphous phase with a higher crystallization temperature upon doping for the data retention process \cite{Andrzej,Urszula2014}. The crystallization temperature for GeTe is $\approx$ 180$^{\circ}$C (453 K)  \cite{Mantovan, Fallica}. GeTe has also been implemented with a superlattice configuration as GeTe-Sb$_{2}$Te$_{3}$ that are extensively used for their application in optical as well as PCM storage devices \cite{Boschker}. Further, the interface between the GeTe and Sb$_{2}$Te$_{3}$ in the superlattice configuration is found to control the phase transition, accompanied by a reduced entropy loss, which helps in making fast and efficient PCMs \cite{Simpson}. Doped GeTe with either N or C has been studied as a way to postpone the phase change for high temperature applications \cite{Andrzej,Fallica}. The effect of doping is also found to reduce the thermal conductivity ($\kappa$) of GeTe significantly \cite{Fallica}. Thermal conductivity ($\kappa$) serves as a crucial parameter for PCM operations as heat dissipation, localization and transport can significantly affect the SET/RESET processes and can therefore considerably influence the performance of PCMs in terms of cyclability, switching time and data retention.

While considerable amount of investigations have been reported on the electronic transport properties of GeTe \cite{Hardik2018,Levin2013}, very few reports on the thermal conductivity of GeTe starting from room temperature to high-temperature range are found to exist in the literature. P. Nath et al. \cite{Nath} characterized the thermal properties of thick GeTe films in both crystalline and amorphous phases. While amorphous phase showed negligible electronic
contribution in the thermal conductivity ($\kappa$), the contribution in crystalline phases were found to be nearly 25$\%$ of the measured value of $\kappa$ \cite{Nath} at the room temperature. E.M. Levin et al. \cite{Levin2013} observed high thermal conductivity of GeTe at 300 K and 720 K which they attributed mostly to free charge carriers. Non-equilibrium molecular dynamics simulation (NEMD) studies by Davide Campi et al. \cite{Campi2015} showed that a 3$\%$ of Ge vacancies effectively reduce the bulk lattice thermal conductivity of crystalline GeTe from 3.2 Wm$^{-1}$K$^{-1}$ to 1.38 Wm$^{-1}$K$^{-1}$ at 300 K, justifying a large spread of the experimentally measured thermal conductivities. First principles calculations \cite{Campi2017} also revealed this large variability of experimentally measured bulk thermal conductivity due to the presence of Ge vacancies. Very recently, frequency domain thermoreflectance study \cite{Ronald2019} was carried out for determining thermal conductivity in GeTe thin films as a function of film thickness for both amorphous and crystalline phases.

Though the aforementioned studies dealt with the thermal transport of GeTe in a broader sense, there are plenty of open questions that still remain. A thorough and systematic understanding of the thermal conductivity of GeTe as a function of temperatures, ranging from room temperature to a temperature that is higher than the crystallization temperature, in the context of different scattering processes involved, is one amongst them. This systematic understanding involves multiple approaches: (a) An accurate experimental estimation of the thermal conductivity of GeTe thin films by taking into account the significant contributions of thermal boundary resistances at the boundaries of the GeTe layer. (b) A consistent and thorough theoretical investigation of the temperature variation of thermal conductivity starting from first-principles density functional theory as well as solving the linearized Boltzmann transport equations (LBTE) and (c) Using simple phenomenological models such as Slack model, which exhibit closed-form solutions that can easily identify the underlying physical mechanism involved in the thermal transport. As revealed in the study by E. Bosoni et al. \cite{Bosoni}, thermal conductivity ($\kappa$) calculated from the Slack model \cite{Morelli2006}, was indeed found in good agreement with the experimentally measured value of $\kappa$ at room temperature.  

In this article, we investigate the thermal conductivity of GeTe films from room temperature up to 230$^{\circ}$C (503 K) starting from the amorphous state. The contribution of the film thermal conductivity from the thermal resistance at the interfaces between the GeTe film and lower and upper layers are clearly discriminated in the whole temperature range. The modulated photothermal radiometry (MPTR) is employed as the experimental technique for the study. The Levenberg-Marquardt (LM) technique is used in order to identify the essential parameters from phase measurements and a model that simulates the phase within the experimental configuration. In a stand-alone exercise of theoretical understanding, thermal conductivity of GeTe is calculated starting from first-principles density functional theory (DFT) coupled with solving linearized Boltzmann transport equations (LBTE) by both direct method and relaxation time approach. In order to explain the measured change in $\kappa\left(T\right)$, distinct contributions coming from various scattering mechanisms are understood. Further, phenomenological models by Slack et al. \cite{slack1964} and Cahill et al. \cite{Cahill} are also employed to elucidate the physical mechanisms in the heat transport process. These systematic theoretical and experimental investigations are found to provide significant clarity and insight in understanding the variation of thermal conductivity of GeTe for a wide range of temperature. This work is organized as follows: Section \ref{section:expt} deals with experimental measurements using MPTR method. Computational details involving first principles and thermal conductivity calculations are described in Section \ref{section:computational}. Section \ref{section:theory} presents the theoretical results followed by summary and conclusions in Section \ref{section:summary}. 

\section{Experimental results}{\label{section:expt}}

Amorphous GeTe films, with thicknesses of 200, 300 and 400 nm, are deposited by magnetron sputtering in an Ar atmosphere on 200 mm silicon wafers covered by a 500 nm thick SiO$_2$ top layer. The thicknesses of the deposited films as well as their homogeneity are controlled
by X-Ray Reflectivity (XRR). The detailed description of the MPTR setup had been presented elsewhere \cite{Andrzej}. The main principle consists of front face periodic heating of the studied sample by a laser source. Since the GeTe layer is not opaque at the laser wavelength (1064 nm) and in order to prevent oxidation or evaporation of the GeTe at high temperature, a 100 nm thick platinum (Pt) layer is deposited by sputtering in order to act as an optical to thermal transducer. The periodic heat flux $\phi\left(\omega\right)$ is thus absorbed by the Pt layer due to its high extinction coefficient at the laser wavelength, and the optical source is then transformed into heat. On the other hand, thanks to its high thermal conductivity and low thickness, the Pt layer is assumed to be isothermal for the frequency range swept during the experiment. The thermal response of the sample at the location of the heating area by the laser is measured using an infrared detector. As the temperature change is low at the heated area, the linearity assumption of heat transfer is fulfilled and the emitted infrared radiation from the sample surface is linearly proportional to the temperature at the heated area. A lock-in amplifier is used to extract the amplitude and the phase from the signal of the IR detector as a function of the frequency. The thermal properties are thus obtained by fitting of the experimental phase by means of a thermal model which allows to describe the heat transfer within the sample. According to the film thickness and modulation frequency, the transient behaviour fulfils the Fourier regime of heat conduction. Since the heated area is much larger (laser spot of $\sim$ 2 mm in diameter) than the film thickness, the one-dimensional heat transfer is considered. The explored frequency range in our experiment is 1-5 kHz. At angular frequency $\omega$, the phase is defined as $\psi\left(\omega\right)=\arg\left[Z\left(\omega\right)\right]=\arctan\left(\mathrm{Im\left(Z\left(\omega\right)\right)}/\mathrm{Re}\left(Z\left(\omega\right)\right)\right)$, where the transfer function $Z\left(\omega\right)$ denotes the ratio between the periodic temperature $\theta\left(\omega\right)$ at the heated area and $\phi\left(\omega\right)$ as:

\begin{equation}
Z\left(\omega\right)=\frac{\theta\left(\omega\right)}{\phi\left(\omega\right)}=\frac{B}{D}\label{eq Zap}
\end{equation}

Parameters $B$ and $D$ are calculated from the quadrupoles formalism \cite{2000-Maillet} as:

\begin{equation}
%\resizebox{1.0\hsize}{!}{
\begin{bmatrix}
A & B \\
C & D
\end{bmatrix}
= \begin{bmatrix}
A_{GeTe} & B_{GeTe} \\
C_{GeTe} & D_{GeTe}
\end{bmatrix}
\begin{bmatrix}
A_{SiO_{2}} & B_{SiO_{2}} \\
C_{SiO_{2}} & D_{SiO_{2}}
\end{bmatrix}
\begin{bmatrix}
A_{Si} & B_{Si} \\
C_{Si} & D_{Si}
\end{bmatrix}
\label{eq quad}
%}
\end{equation}

Where

\begin{eqnarray}
A_{j} & = & 1+\exp\left(-2\,\gamma_{i}\,e_{i}\right);\,B_{j}=\frac{\left(1+\exp\left(-2\,\gamma_{i}\,e_{i}\right)\right)}{\gamma_{i}\kappa_{i}^{*}}\label{eq cQuads}\\
C_{j} & = & \left(1+\exp\left(-2\,\gamma_{i}\,e_{i}\right)\right)\,\gamma_{i}\,\kappa_{i}^{*} ; \,D_{j}=A_{j}
\end{eqnarray}

\begin{figure}[H]
\centering
\includegraphics[width=0.5\textwidth]{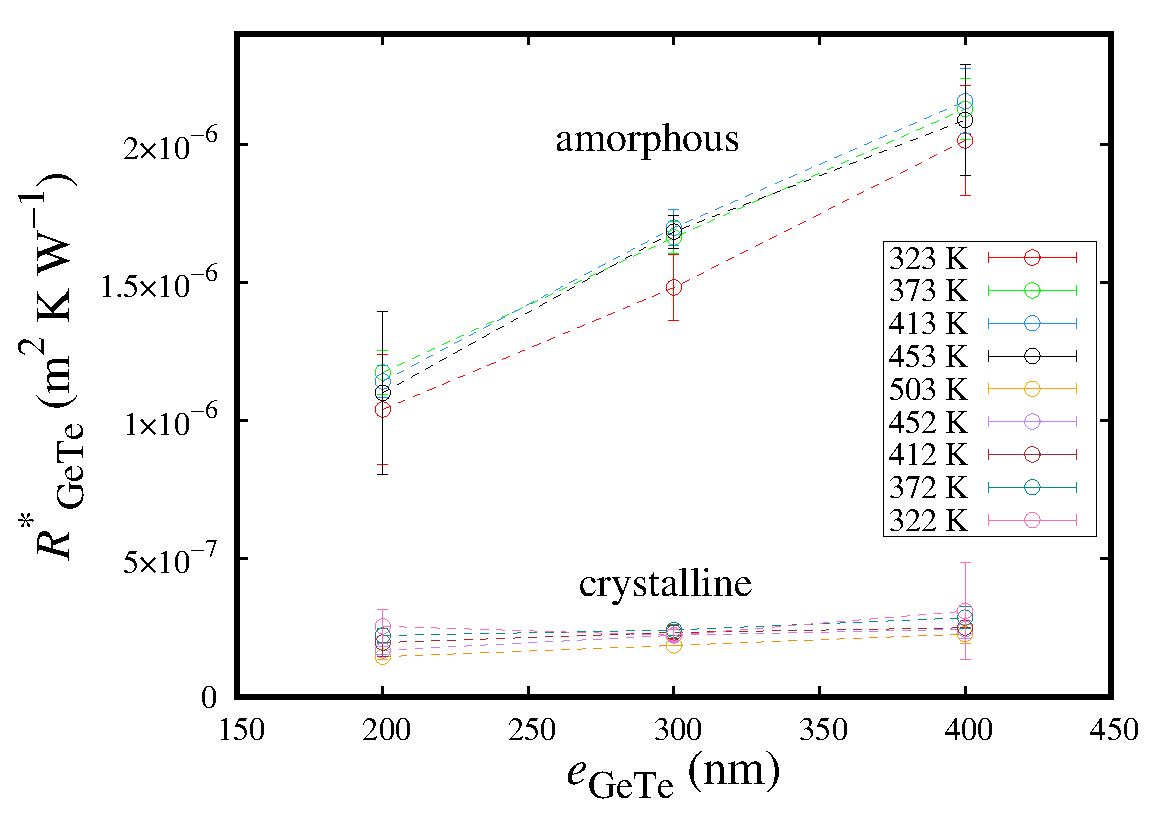}
\caption{\label{fig MPTR-results} Variation of thermal resistance ($R^{*}_{GeTe}$) of GeTe thin film as a function of film thickness ($e_{GeTe}$) for different temperatures ranging from amorphous to crystalline phase change.}
\end{figure}

With $\gamma_{i}=\sqrt{j\,\omega/a_{i}^{*}}$, where $a_{i}^{*}$($=\kappa_{i}^{*}/C_{p_{i}}$), $\kappa_{i}^{*}$, $C_{p_{i}}$ and $e_{i}$ are the effective thermal diffusivity, effective thermal conductivity, specific heat per unit volume and thickness of layer $i$ respectively. For the SiO$_{2}$ and Si layers, the effective thermal conductivity is equal to the real thermal conductivity, i.e., $\kappa_{SiO_{2}}^{*}=\kappa_{SiO_{2}}$ and $\kappa_{Si}^{*}=\kappa_{Si}$. On the other hand, the effective thermal resistance for GeTe ($R_{GeTe}^{*}$) accounts with the intrinsic thermal conductivity $\kappa_{GeTe}$ of the GeTe layer and the thermal resistance at the two interfaces with Pt and SiO$_{2}$ as:

\begin{equation}
R_{GeTe}^{*} = \frac{e_{GeTe}}{\kappa_{GeTe}^{*}}=\frac{e_{GeTe}}{\kappa_{GeTe}}+\underbrace{R_{Pt-GeTe}+R_{GeTe-SiO_{2}}}_{R_i}\label{eq_thermal_resistance}
\end{equation}
where $R_{i}$ is the total interfacial thermal resistance. It must be precisely mentioned here that the GeTe layer cannot be considered as thermally resistive for the highest frequency values and especially for the high thickness of the layer. This is the reason why we consider heat diffusion within the quadrupole model. This can be easily demonstrated by calculating the Fourier related quantity $\sqrt{2\,a_{GeTe}/\omega}$, using the known value of $a_{GeTe}$ at room temperature and comparing it to the thickness $e_{GeTe}$. Considering the measured value $Y_{\phi}\left(\omega_{i}\right)$ of the phase at different frequencies $\omega_{i}\,\left(i=1..N\right)$, the value of $\kappa_{GeTe}^{*}$ is estimated by minimizing the objective function $J=\left\Vert \mathbf{Y_{\phi}}-\mathbf{\mathbf{\Psi}}\right\Vert _{2}$, where $\mathbf{Y_{\phi}=\mathit{Y_{\phi}\left(\omega_{i}\right)_{i=1..N}}}$ and $\mathbf{\Psi}=\mathbf{\mathbf{\psi}}\left(\omega_{i}\right)_{i=1..N}$ are vectors with length $N$, related respectively to the measured and simulated phase at all the investigated frequencies. This minimization is achieved by implementing the Lavenberg-Marquardt (LM) algorithm \cite{ASTER2019ix}. Then, the effective thermal resistance $R_{GeTe}^{*}$ of GeTe thin film is plotted as a function of film thicknesses ($e_{GeTe}$) for different temperatures as shown in Fig \ref{fig MPTR-results}.  

A linear regression $R_{GeTe}^{*}=\alpha\,e_{GeTe}+\beta=e_{GeTe}/\kappa_{GeTe}+R_{i}$ is found that allows to extract\\
\\

$\kappa_{GeTe}=1/\alpha$ from the slope (shown in Fig \ref{fig MPTR-results_kappa}.(b)) and the sum of the two interfacial resistances $R_{i}=\beta$ by extrapolation to $e_{GeTe}=0$ (shown in Fig \ref{fig MPTR-results_kappa}.(a)). The standard deviation of $\kappa_{GeTe}^{*}$ is calculated from the covariance matrix at the end of the iterative minimization process as: $\sigma\left(\kappa_{GeTe}^{*}\left|Y\right.\right)^{2}\sim\mathbf{cov}\left(\mathbf{\Theta}\right)\mathbf{E}/\sqrt{N}$ where the covariance matrix is: $\mathbf{cov}\left(\mathbf{\Theta}\right)=\left(\mathbf{S}^{T}\mathbf{S}\right)^{-1}$
with vector $\mathbf{S}=\left[S_{Q}\left(\alpha_{i}\right)\right]_{N}$
with $S_{Q}\left(\alpha_{i}\right)=\left[\partial\mathbf{\mathbf{\psi}}\left(\omega_{i}\right)_{i=1,N}/\partial\kappa_{GeTe}^{*}\right]_{\kappa_{GeTe}^{*}=\hat{\kappa}_{GeTe}^{*}}$
denoting the sensitivity function of the phase according to $\kappa_{GeTe}^{*}$
calculated for $\kappa_{GeTe}^{*}=\hat{\kappa}_{GeTe}^{*}$ where
$\hat{\kappa}_{GeTe}^{*}$ is the optimal value for $\kappa_{GeTe}^{*}$.
Finally the residual vector is $E=\mathbf{Y_{\phi}}-\mathbf{\Psi}\left(\kappa_{GeTe}^{*}=\hat{\kappa}_{GeTe}^{*}\right)$. The standard deviation on $R_{GeTe}^{*}$ is obtained  starting from $\sigma(\kappa_{GeTe}^{*})$ by application of the law of propagation of uncertainties. Finally, the standard deviations on $\kappa_{GeTe}$ and $R_{i}$ are expressed according to residual variance of linear fitting on $R_{GeTe}^{*}$ points. We mention here that the grain size of GeTe at the time of the crystallization is found to be 40 nm as reported in our earlier work \cite{Andrzej} and it increases while increasing the annealed temperature (Fig 4 in \cite{Andrzej}). Since the MPTR investigates a very large area, it is thus expected to measure the average thermal conductivity that is given by $\kappa_L^{av}$ + $\kappa_{el}$, where  $\kappa_L^{av}$ = $\frac{2}{3}$$\kappa_x$ + $\frac{1}{3}$$\kappa_z$, where, $\kappa_x$ and $\kappa_z$ stand for the lattice thermal conductivities along hexagonal a and c axes respectively.

The phase change occurs well within the range of the expected temperature ($\sim$ 180$^{\circ}$C = 453 K). We note that $R_{i}$ for amorphous state is difficult to estimate because of the very low values of $\kappa_{GeTe}$. For crystalline phase, we retrieve a consistent behavior with the value of $R_{i}$ being increased as the temperature is lowered (Fig \ref{fig MPTR-results_kappa}.(a)). In the high temperature regime, diffuse mismatch model (DMM) has been found to describe $R_{i}$ quite satisfactorily for crystalline solids \cite{Reifenberg}. According to DMM model, 
\onecolumngrid
\begin{widetext}
\vspace{-0.4cm}
\begin{figure}[H]
\centering
\includegraphics[width=1.0\textwidth]{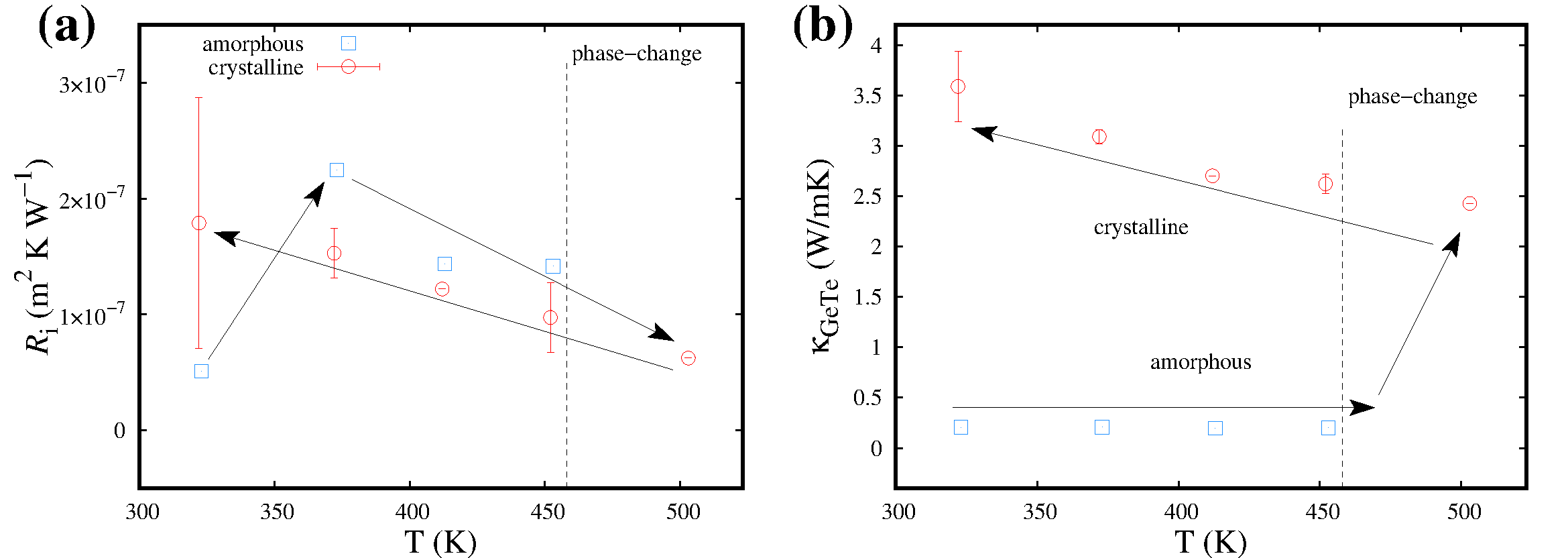}
\caption{\label{fig MPTR-results_kappa} (a) Variation of total interfacial thermal resistance $R_{i}=R_{Pt-GeTe}+R_{GeTe-SiO_{2}}$ with temperature. Phase change occurs around 180 $^{\circ}$C = 453 K, which is shown via dotted line. (b) Experimentally measured thermal conductivity $\kappa_{GeTe}\left(T\right)$ of the GeTe thin film as a function of temperature. The rightward arrows denote the forward cycle exhibiting phase change from amorphous to crystalline phase while the leftward arrow defines the backward cycle where GeTe exists in crystalline phase.}
\end{figure}
\end{widetext}
asymptotic behavior of $R_{i}$ at high temperatures depends inversely on the heat capacity as \cite{Reifenberg}
\begin{equation}
    R_{i} = \left(\frac{\sum_{j} c_{2, j}^{-2}}{12 \left(\sum_{j} c_{1, j}^{-2} + \sum_{j} c_{2, j}^{-2} \right)} \sum_{j} c_{1, j}\right)^{-1} \frac{1}{C_{1}\left(T\right)}
\end{equation} 
where, $C_{1}(T)$ is the heat capacity of material 1 at $T$ and $c_{l,j}$ is the velocity of phonon mode $j$ in material $l$. Here all the parameters except $C_{1}(T)$ are temperature independent. Since $C_{1}(T)$ increases with temperature, the above relation shows that $R_i$ decreases with increasing temperature. In Fig \ref{fig MPTR-results_kappa}.(b), we observe a monotonically decreasing trend of $\kappa_{GeTe}$ as a function of T. Indeed, at high temperatures (T $\gg$ $\Theta_D$), where $\Theta_D$ is Debye temperature, umklapp scattering is the dominating phonon-phonon scattering process associated with high momentum change in phonon-phonon collisions \cite{Kittel86}. Glen A. Slack et al.\cite{slack1964} approximated umklapp relaxation time as $\tau_U^{-1}$ = $AT\omega^{2}exp\left(-\Theta_{D}/3T \right)$ which becomes $\tau_U^{-1}$ = $AT\omega^{2}$ when T $\gg$ $\Theta_D$. This relaxation time estimation leads to $\kappa$ $\propto$ 1/$\tau_{U}^{-1}$ $\propto$ 1/$T$. 

\section{Computational details}{\label{section:computational}}
%\subsection{Phonon Density of states from density functional theory}
Phonon density of states (PDOS) of crystalline GeTe (space group R3m) has been calculated employing the density functional perturbation theory (DFPT) \cite{Baroni} using the QUANTUM-ESPRESSO \cite{qe} suite of programs. As the first step, self-consistent calculations, within the framework of density functional theory (DFT), are carried out to compute the total ground state energy of the crystalline R3m-GeTe. For this purpose, Perdew-Burke-Ernzerhof (PBE) \cite{PBE} generalized gradient approximation (GGA) is used as the exchange-correlation functional. The spin-orbit interaction has been ignored due to its negligible effects on the vibrational features of GeTe as mentioned in literature \cite{Shaltaf2009,Campi2017}. Electron-ion interactions are represented by pseudopotentials using the framework of projector-augmented-wave (PAW) method \cite{PAW}. The Kohn-Sham (KS) orbitals are expanded in a plane-wave (PW) basis with a kinetic cutoff of 60 Ry and a charge density cutoff of 240 Ry as prescribed by the pseudopotentials of Ge and Te. The Brillouin zone integration for self consistent electron density calculations are performed using a 12$\times$12$\times$12 Monkhorst-Pack (MP) \cite{MP} k-point grid.

For phonon calculations, a hexagonal 2$\times$2$\times$1 supercell, consisting of 24 atoms, is used. A representative 
\begin{figure}[H]
    \centering
    \includegraphics[width=0.55\textwidth]{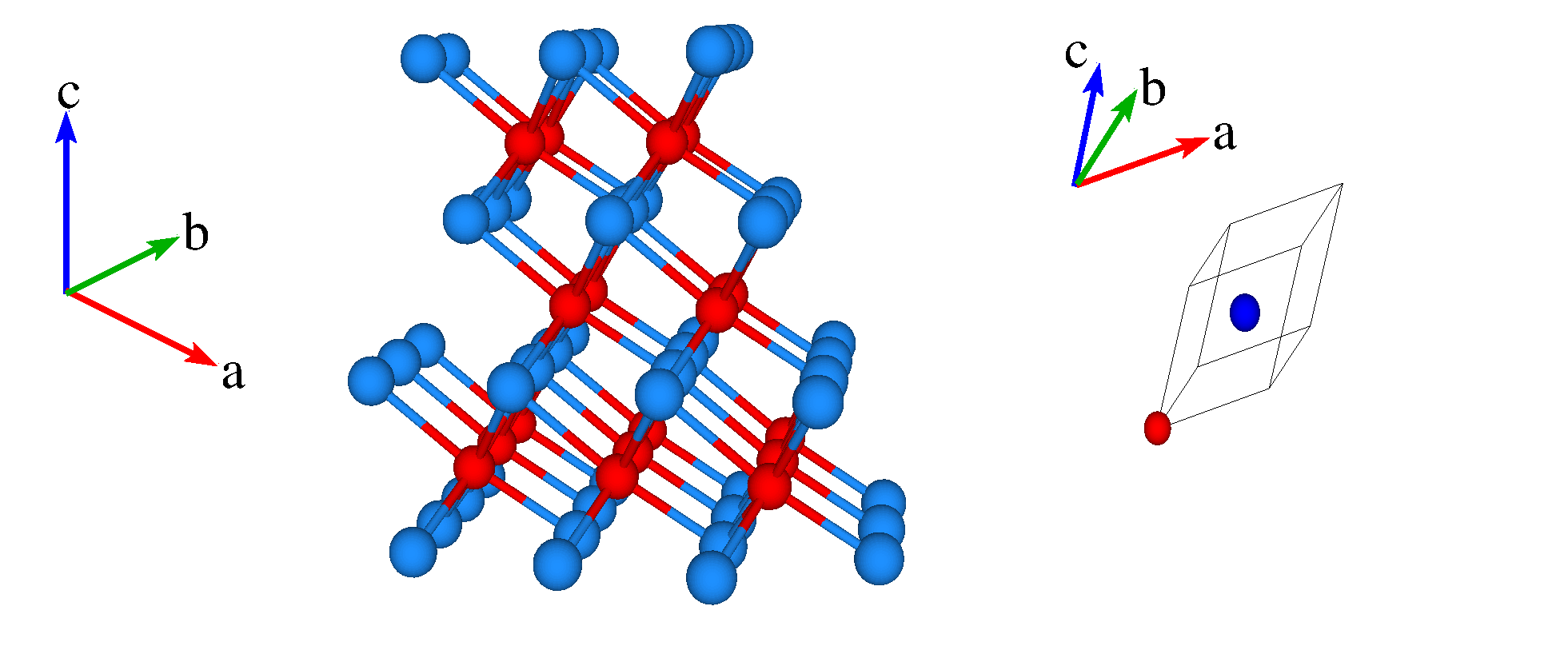}
    \caption{2$\times$2$\times$1 supercell of GeTe crystal (R3m) and its trigonal primitive cell. The structure of GeTe can be seen as a stacking of bilayers along c axis of hexagonal unit cell as mentioned in \cite{Campi2015,Campi2017}. Here red spheres denote Ge atoms and blue spheres denote Te atoms. This representation is realized through VESTA \cite{vesta}. Here, Ge atoms at the boundary take into account the additional Te atoms outside the cell.}
    \label{fig:1}
\end{figure}
structure of crystalline GeTe as stacking bilayers is shown in Fig \ref{fig:1}. To study the phonon density of states, linear response theory is applied via DFPT, to the Kohn-Sham equations to solve the electronic charge density ($\rho_n$) under small perturbations. As the force constants are connected to the derivatives of $\rho_n$ with respect to atomic displacements, harmonic force constants are calculated by diagonalizing the dynamical matrix in reciprocal space. Phonon density of states (PDOS) are then evaluated by the inverse Fourier transform of the interatomic force constants (IFC) to real space from that of the dynamical  matrices, using a uniform  5$\times$5$\times$5 grid of $\textbf{q}$-vectors.

%\subsection{Electronic thermal conductivity}
The thermal conductivity of GeTe can be separated into two distinct contributions, one from electronic transport and the other from the phonon transport or the lattice contribution, such that $\kappa$ = $\kappa_{el}$ + $\kappa_{L}$, where $\kappa_{L}$ is the lattice thermal conductivity and $\kappa_{el}$ is the electronic thermal conductivity. $\kappa_{el}$ has been obtained from first-principles calculations by solving semi-classical Boltzmann transport equation (BTE) for electrons. Constant relaxation time approximation (CRTA) and rigid band approximation (RBA) are employed as implemented in BoltzTraP code \cite{BoltzTraP}. The energy projected conductivity tensor is calculated using:
\begin{equation}
    \sigma_{\alpha\beta}(\epsilon) = \frac{1}{N} \sum_{i, \textbf{k}} \sigma_{\alpha\beta} (i, \textbf{k}) \frac{\delta(\epsilon - \epsilon_{i,\textbf{k}})}{d\epsilon}
\end{equation}
Therefore, the transport tensors, or more specifically the electrical conductivity tensor in this study, can be obtained from 
\begin{equation}
    \sigma_{\alpha\beta} (T; \mu) = \frac{1}{\Omega} \int \sigma_{\alpha\beta}(\epsilon)\left[-\frac{\partial f_{\mu}(T;\epsilon)}{\partial \epsilon}\right]d\epsilon 
\end{equation}
where, N is the number of $\textbf{k}$-points sampled, $i$ is the band index, $\epsilon_{i,\textbf{k}}$ are band energies, $\Omega$ is the volume of unit cell, $f_\mu$ is the Fermi distribution function and $\mu$ is the chemical potential. The code computes the Fermi integrals and returns the transport coefficients for different temperature and Fermi levels.

%\subsection{Lattice thermal conductivity}
For getting lattice thermal conductivity $\kappa_L$, linearized phonon Boltzmann transport equation (LBTE) is solved using both direct method introduced by L. Chaput et al. \cite{Chaput} as well as the single mode relaxation time approximation or relaxation time approximation (RTA) employing PHONO3PY \cite{Togo} software package. Initially, the supercell approach with finite displacement of 0.03 \AA{} is applied to calculate the harmonic (second order) and the anharmonic (third order) force constants, given by 
\begin{equation}
    \Phi_{\alpha \beta} (l\kappa, l'\kappa') = \frac{\partial^2 \Phi}{\partial u_{\alpha} (l\kappa)\partial u_{\beta} (l'\kappa')}
\end{equation}
and
\begin{equation}
    \Phi_{\alpha \beta \gamma} (l\kappa, l'\kappa', l''\kappa'') = \frac{\partial^3 \Phi}{\partial u_{\alpha} (l\kappa)\partial u_{\beta} (l'\kappa') \partial u_{\gamma} (l''\kappa'')}
\end{equation}
respectively. First principles calculations using QUANTUM-ESPRESSO \cite{qe} are implemented to calculate the forces acting on atoms in supercells. Using finite difference method, harmonic force constants are approximated as \cite{Togo} 
\begin{equation}
   \Phi_{\alpha \beta} (l\kappa, l'\kappa') \simeq - \frac{F_{\beta} [l'\kappa'; \textbf{u} (l \kappa)]}{u_{\alpha} (l \kappa)} 
\end{equation}
where \textbf{F}[$l'$$\kappa'$; \textbf{u}($l$$\kappa$)] is atomic force computed at \textbf{r}($l'$ $\kappa'$) with an atomic displacement \textbf{u}($l\kappa$) in a supercell. Similarly, anharmonic force constants are obtained using\cite{Togo} 
\begin{equation}
   \Phi_{\alpha \beta \gamma} (l\kappa, l'\kappa', l''\kappa'') \simeq - \frac{F_{\gamma} [l''\kappa''; \textbf{u} (l \kappa), \textbf{u} (l' \kappa')]}{u_{\alpha} (l \kappa)u_{\beta} (l' \kappa')} 
\end{equation}
where \textbf{F}[$l''$$\kappa''$; \textbf{u}($l$$\kappa$), \textbf{u}($l'$ $\kappa'$)] is atomic force computed at \textbf{r}($l''$ $\kappa''$) with a pair of atomic displacements \textbf{u}($l\kappa$) and \textbf{u}($l'\kappa'$) in a supercell. These two sets of linear equations are solved using Moore-Penrose pseudoinverse as is implemented in PHONO3PY \cite{Togo}.

We use a 2$\times$2$\times$2 supercell of GeTe for our first-principles calculations of anharmonic force constants. Using the supercell and finite displacement approach, 228 supercells are obtained, having different pairs of displaced atoms, for the calculations for the anharmonic force constants. A larger 3$\times$3$\times$3 supercell is employed for calculating the harmonic force constants.  For all the supercell force calculations, the reciprocal space is sampled using a 3$\times$3$\times$3 k-sampling MP mesh shifted by a half grid distances along all three directions from $\Gamma$- point. The total energy convergence threshold has been kept at 10$^{-10}$ a.u. for supercell calculations. For lattice thermal conductivity calculations employing both the direct solution of LBTE and that of the RTA, $\textbf{q}$-mesh of 24$\times$24$\times$24 are used. The imaginary part of self-energy have been calculated using tetrahedron method from which phonon lifetimes are obtained . 

%\section{Results and Discussions}

\section{Theoretical results and Discussions} {\label{section:theory}}
\subsection{Phonon density of states}

The structural parameters are optimized via DFT calculations and the optimized lattice parameter (a = 4.23 \AA) and unit cell volume (56.26 \AA$^3$) of GeTe, are found to be quite consistent with the values presented in literature \cite{Campi2017,Jain2013}. It is quite well established that at normal conditions GeTe crystallizes in trigonal phase (space group R3m) with 2 atoms per unit cell. This structure gives rise to a 3+3 coordination of Ge with three short stronger intrabilayer bonds and three long weaker interbilayer bonds \cite{Campi2015,Campi2017}. The bond lengths (shorter bonds = 2.85 \AA, longer bonds = 3.25 \AA) are also found to be consistent with the studies done by Davide Campi et al. \cite{Campi2017}.

To investigate the effect of phonons in the heat transfer processes, we study the phonon density of states and the  
\onecolumngrid
\begin{widetext}
\begin{figure}[H]
\centering
\includegraphics[width=1.0\textwidth]{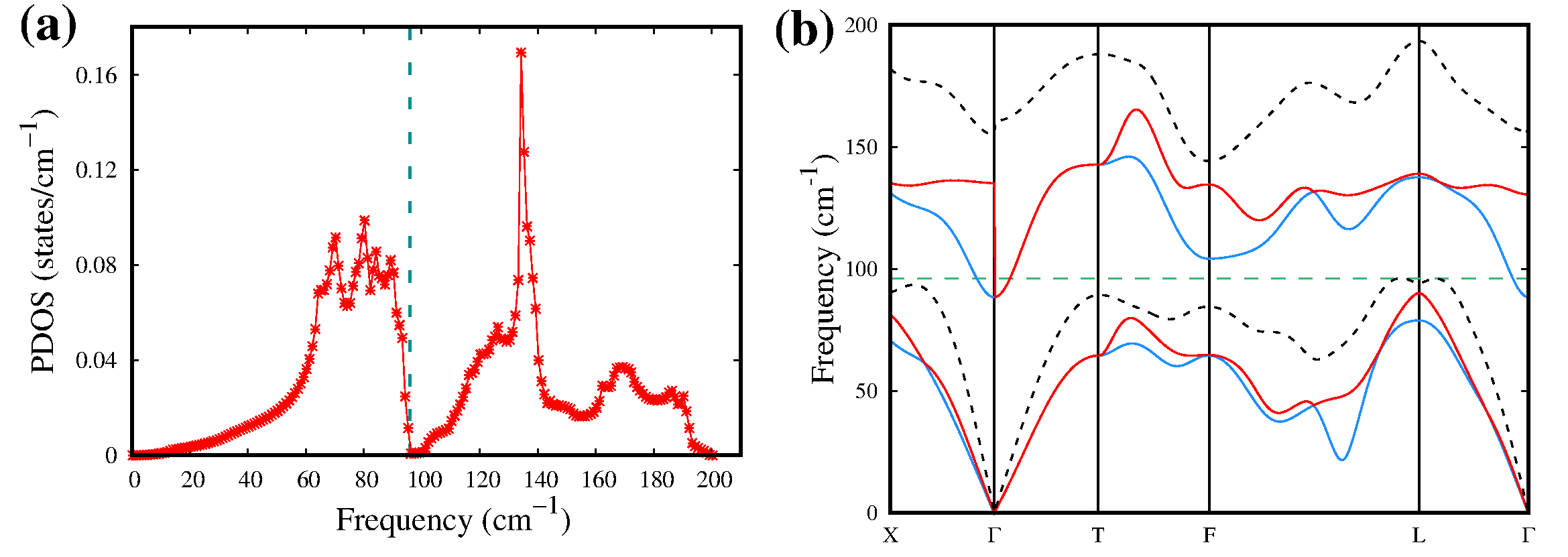}
\caption{\label{PDOS} (a) Calculated  phonon density of states (PDOS) for GeTe crystal (R3m). The dotted line denotes the separation frequency between acoustic and optical modes ($\sim$ 96 cm$^{-1}$ = 2.88 THz) which is consistent with previous studies (see text). (b) Phonon dispersion relation of the rhombohedral (R3m) GeTe. Transverse and longitudinal phonon modes are denoted via solid and dashed lines, respectively. The approximate separation frequency between acoustic and optical modes ($\sim$ 96 cm$^{-1}$) are shown via green dashed line.}
\end{figure}
\vspace{-0.5cm}
\end{widetext}
dispersion relation of crystalline rhombohedral (R3m) GeTe. Figure \ref{PDOS}.(a) and (b) show the phonon density of states and the phonon dispersion relation of undoped crystalline (R3m) GeTe, calculated at ground state. The dashed line in Fig \ref{PDOS}.(a) serves as a separator between the acoustic and optical contribution of phonons. The acoustic phonons extend from 0 to 96 cm$^{-1}$ (= 2.88 THz) in the frequency domain which is consistent with the observations of Urszula D. Wdowik et al. \cite{Urszula2014}. The frequencies of the two most prominent peaks in PDOS corresponding to acoustic ($\sim$ 80 cm$^{-1}$ = 2.40 THz) and optical phonons ($\sim$ 140 cm$^{-1}$ = 4.20 THz) respectively, are also found to be close to the values observed by Kwangsik Jeong et al. \cite{Jeong2017}. Phonon dispersion relation (Fig \ref{PDOS}.(b)) along the high symmetry direction in the Brillouin zone (BZ) also shows similar trends to that of the earlier works \cite{Urszula2014, Bosoni}. We observe the signature of LO-TO splitting as the discontinuities of the phonon dispersion at the $\Gamma$ point arising from the long range Coulomb interactions \cite{Urszula2014, Bosoni}. The approximate separator between acoustic and optical modes at 96 cm$^{-1}$ is also shown by a horizontal dashed line in the phonon dispersion relation (Fig \ref{PDOS}.(b)). Except a small and negligible contribution of transverse optical modes at the $\Gamma$ point, all frequencies $<$ 96 cm$^{-1}$ contribute to the acoustic modes.

\subsection{Electronic thermal conductivity}

 The calculation of the electronic thermal conductivity $\kappa_{el}$ rests generally on the use of the Wiedemann-Franz law $\kappa_{el} = L\sigma_{e}T$ where $L$ is the Lorenz number, $T$ is the temperature and $\sigma_{e}$ is the electrical conductivity. Crystalline GeTe is a $\textit{p}$-type degenerate semiconductor \cite{Bahl,Nath} with a high hole concentration and the Fermi level lying well inside the valence band that holds the charge carriers. Therefore it behaves similar to a metal. In that case, the value for $L=\pi^{2}\left(k_{B}/e\right)^{2}/3=2.44\times10^{-8}\,\mathrm{V^{2}.K^{-2}}$ ($k_{B}$ is the Boltzmann constant and $e$ is the electron charge) had generally been employed in most of the research works \cite{Nath}. However, for highly doped materials at temperature higher than the Debye temperature $\Theta_{D}$, the Hall mobility depends on the temperature as $\mu_{H}\propto T^{-(3/2)+r}$ , where $r$ is the scattering mechanism
parameter, and decreases as \ensuremath{\eta} decreases with increasing temperature. Therefore, it is shown that the Lorenz number is given by \cite{Goldsmid2010, Gelbstein2010}:
\begin{widetext}
\begin{equation}
L=\left(\frac{k_{B}}{e}\right)^{2}\left[\frac{\left(r+\frac{7}{2}\right)\left(r+\frac{3}{2}\right)\,F_{r+5/2}\left(\eta\right)\,F_{r+1/2}\left(\eta\right)-\left(r+\frac{5}{2}\right)^{2}\,F_{r+3/2}^{2}\left(\eta\right)}{\left(r+\frac{3}{2}\right)^{2}\,F_{r+1/2}^{2}\left(\eta\right)}\right]\label{eq Lorentz}
\end{equation}
\end{widetext}
where $\eta=\left(E_{F}-E_{V}\right)/k_{B}T$ is the reduced Fermi energy for $p$-type semiconductors and $r=-1/2$ when scattering by acoustic phonons is dominant \cite{Gelbstein2010}. The Fermi integral is defined as: $F_{n}\left(\eta\right)=\int_{0}^{\infty}\left(x^{n}/\left(1+e^{x-\eta}\right)\right)\textit{d}x$. 
\begin{table}[H]%The best place to locate the table environment is directly after its first reference in text
\caption{\label{tab:table_k_el}%
Electronic thermal conductivity ($\kappa_{el}$) of the crystalline R3m-GeTe is presented (column 4) as a function of temperature (T) using Wiedemann-Franz law. The variation of electrical conductivity ($\sigma_e$) and the Lorenz number (L) with temperature are also shown in column 2 and column 3 respectively.}
\begin{ruledtabular}
\begin{tabular}{cccc}
T (K) & $\sigma_e$ ($\Omega^{-1}m^{-1}$) & L (W$\Omega$ $K^{-2}$) & $\kappa_{el}$ = $L\sigma_{e}T$ (W/mK) \\
\colrule
322 & 9.78$\times10^4$ & 2.40$\times10^{-8}$ & 0.76 \\
372 & 9.35$\times10^4$ & 2.33$\times10^{-8}$ & 0.81 \\
412 & 9.14$\times10^4$ & 2.27$\times10^{-8}$ & 0.85 \\
452 & 8.84$\times10^4$ & 2.20$\times10^{-8}$ & 0.88 \\
503 & 8.45$\times10^4$ & 2.13$\times10^{-8}$ & 0.91 \\
\end{tabular}
\end{ruledtabular}
\end{table}
Following the method adopted by Gelbstein et al. \cite{Gelbstein2010}, E. M. Levin and co authors \cite{Levin2013} obtained that $L$ for crystalline GeTe varies between $2.4\times10^{-8}\,\mathrm{V^{2}.K^{-2}}$ at 320 K and $1.8\times10^{-8}\,\mathrm{V^{2}.K^{-2}}$ at 720 K. Using this method, we calculate the variation of $L$ for GeTe as a function of temperature within the investigated range in the present study and we report the results in Table \ref{tab:table_k_el}.

The electrical resistivity of GeTe films has been measured experimentally using the Van der Pauw technique and is found to be $\rho_{e}=1/\sigma_{e}=\left[8.5\pm2\right]\times10^{-6}\,\Omega.\mathrm{m}$ at the room temperature (300 K), which corresponds for the hole concentration to be 6.24 $\times$ 10$^{19}$ cm$^{-3}$. The constant relaxation time approximation (CRTA) with a constant electronic relaxation time  of 10$^{-14}$s is used following the work of S. K. Bahl et al. \cite{Bahl}. After identifying the hole concentration, we compute the electrical conductivity ($\sigma_e$) by using DFT and solving the Boltzmann transport equation (BTE) for electrons, as implemented in the BoltzTaP code \cite{BoltzTraP} for the given hole concentration at different temperatures. Considering the values for $L\left(T\right)$, the calculated electronic thermal conductivity varies linearly from 0.76 W/mK at 322 K ($L=2.40\times10^{-8}\,\mathrm{V^{2}.K^{-2}}$ at 322 K) to 0.91 W/mK at 503 K ($L=2.13\times10^{-8}\,\mathrm{V^{2}.K^{-2}}$ at 503 K) (Table \ref{tab:table_k_el}). These results are consistent with the experimental ones by R. Fallica et al. \cite{Fallica}.

\subsection{Lattice thermal conductivity}

In Fig \ref{fig MPTR-results_kappa}.(b), total thermal conductivity of GeTe is seen to manifest a fast change process with gradually increasing temperature, where a phase change from amorphous to crystalline phase is found to occur $\sim$ 180$^{\circ}$C or 453 K. We first focus on the thermal conductivity of the amorphous phase of GeTe. Figure \ref{fig:kappa_expt_theory} shows lower values of $\kappa$ for amorphous phase compared to that of the crystalline phase. The values of $\kappa$ in amorphous phase are also observed to be almost constant throughout the temperature range studied in this work. Theoretically, the minimal thermal conductivity model derived by Cahill et al. \cite{Cahill} allows one to calculate $\kappa$ for the amorphous materials as:
%\begin{widetext}
\begin{equation}
    \textstyle \kappa_{min}\left(T\right)=\left(\frac{\pi}{6}\right)^{1/3}\,k_{B}\,n^{2/3}\,\sum_{i=1}^{3}v_{i}\,\left(\frac{T}{\Theta_{i}}\right)^{2}\int_{0}^{\Theta_{i}/T}\frac{x^{3}\,e^{x}}{\left(e^{x}-1\right)^{2}}\,\mathrm{d}x
\end{equation}
%\end{widetext}
where $n=\left(k_{B}\,\Theta_{D}/\hbar\right)^{-1/3}/\left(6\,\pi^{2}\,c_{s}^{3}\right)$
is the number of phonons per unit volume (which can also be calculated
more rigorously from the phonon DOS apart from using the values of Table \ref{table:parameters_cahill_slack}), $c_s$ is the speed of sound calculated
as $c_{s}^{-3}=\left(v_{L}^{-3}+2\,v_{T}^{-3}\right)/3=1900\,\mathrm{m.s^{-1}}$
\cite{Pereira} and $\Theta_{i}$ is the Debye temperature per
branch. When $T\gg\Theta_{D}$, this relation simplifies as
\begin{equation}
    \kappa_{min} =\frac{1}{2}\left(\pi\,n^{2}/6\right)^{1/3}\,k_{B}\,\left(v_{L}+2\,v_{T}\right).
\end{equation}
Using the required parameter values from Table \ref{table:parameters_cahill_slack}, the minimal thermal conductivity ($\kappa_{min}$) is found to be consistent with the experimental data in amorphous phase  considering the error bars involved in the experimental measurements (Fig \ref{fig:kappa_expt_theory}). The reasonable agreement between the experimental data and $\kappa_{min}$ based on Cahill model indicates that the dominant thermal transport in the amorphous GeTe occurs in short length scales \cite{kaviany_2014} between neighboring vibrating entities owing to the disorder present in it.

We then investigate the phonon contributions to the total thermal conductivity of crystalline GeTe. In order to evaluate the lattice thermal conductivity ($\kappa_L$) through the direct solution of LBTE, the method developed by L. Chaput \cite{Chaput} is adopted. According to this method, lattice thermal conductivity is given as \cite{Chaput} 
\begin{equation}
    \kappa_{\alpha\beta} = \frac{\hbar^2}{4k_{B}T^{2}NV_{0}} \sum_{\lambda\lambda'} \frac{\omega_{\lambda}\upsilon_{\alpha}(\lambda)}{sinh(\frac{\hbar\omega_{\lambda}}{2k_{B}T})}\frac{\omega_{\lambda'}\upsilon_{\beta}(\lambda')}{sinh(\frac{\hbar\omega_{\lambda'}}{2k_{B}T})} (\Omega^{\sim1})_{\lambda\lambda'}
\end{equation}
where, $\Omega^{\sim1}$ is the Moore-Penrose inverse of the collision matrix $\Omega$, given by \cite{Chaput,Togo}
\begin{widetext}
\begin{equation}
    \Omega_{\lambda\lambda'} = \delta_{\lambda\lambda'}/\tau_{\lambda} + \pi/\hbar^{2} \sum_{\lambda''}\mid \Phi_{\lambda\lambda'\lambda''}\mid ^{2}\frac{[\delta(\omega_{\lambda}-\omega_{\lambda'}-\omega_{\lambda''}) + \delta(\omega_{\lambda}+\omega_{\lambda'}-\omega_{\lambda''}) + \delta(\omega_{\lambda}-\omega_{\lambda'}+\omega_{\lambda''})]}{sinh(\frac{\hbar\omega_{\lambda''}}{2k_{B}T})}
\end{equation}
\end{widetext}

Here, $\Phi_{\lambda\lambda'\lambda''}$ denotes the interaction strength between three phonon $\lambda$, $\lambda'$ and $\lambda''$ scattering \cite{Togo}. However, adopting the relaxation time approximation (RTA) in solving LBTE, lattice thermal conductivity tensor $\boldsymbol{\kappa}_L$ can be written in a convenient and closed form as \cite{Giorgia,Togo}
\onecolumngrid
\begin{widetext}
\begin{table}[H]
\vspace{-0.91cm}
\caption{\label{table:parameters_cahill_slack} Theoretical and experimental values of different parameters used for thermal conductivity calculation using Slack \cite{Morelli2006} and Cahill \cite{Cahill} model.}
\begin{ruledtabular}
\begin{tabular}{ccc}
GeTe (R3m)  & Parameter description(s)  & Value(s)\\
\hline 
$V_{0}$ [\AA$^{3}$] & Volume of the elementary cell  & 56.26 \cite{Jain2013},[calculated from DFT]\\
$\rho$ [kg m$^{-3}$] & Density  & 5910 \cite{Jain2013},[calculated from DFT]\\
$M_{Ge}$ [g mol$^{-1}$] & Molar mass of Ge  & 72.63\\
$M_{Te}$ [g mol$^{-1}$] & Molar mass of Te  & 127.6\\
$\Theta_{D}$ [K] & Debye temperature  & 180 \cite{Campi2017}\\
$\upsilon_{L}$ [m s$^{-1}$]  & Phonon group velocity (longitudinal)  & 2500 \cite{Pereira}\\
$\upsilon_{T}$ [m s$^{-1}$]  & Phonon group velocity (Transverse)  & 1750 \cite{Pereira}\\
$G$ & Gruneisen parameter & 1.7 \cite{Bosoni}\\
$E_{F}$ [eV]  & Fermi energy & 7.2552 [calculated from DFT]\\
$N$ [kg$^{-1}$] & number of phonon per unit mass & $5.6723\times10^{24}$ [calculated from PDOS]\\
\end{tabular}
\end{ruledtabular}
\end{table}
\end{widetext}
\begin{equation}{\label{equation_kl}}
    \boldsymbol{\kappa}_L = \frac{1}{NV_0} \sum_{\lambda} C_{\lambda} \textbf{v}_{\lambda} \otimes \textbf{v}_{\lambda}\tau_{\lambda}
\end{equation}
where $N$ is the number of unit cells and $V_0$ is the volume of unit cell.  The phonon modes ($\textbf{q}$, $j$) comprising wave vector $\textbf{q}$ and branch $j$ are denoted with $\lambda$. The modal heat capacity if given by 
\begin{equation}
    C_\lambda = k_{B} \left(\frac{\hbar\omega_{\lambda}}{k_{B}T}\right)^{2} \frac{exp(\hbar\omega_{\lambda}/k_{B}T)}{[exp(\hbar\omega_{\lambda}/k_{B}T) -1]^2}
\end{equation}
Here, $T$ denotes temperature, $\hbar$ is reduced Planck constant and $k_B$ is the Boltzmann constant. $\textbf{v}_\lambda$ and $\tau_\lambda$ represent phonon group velocity and phonon lifetime respectively. We consider three scattering processes, namely normal, umklapp and isotope, denoted by N, U and I respectively, in the theoretical study. For each of these processes, the phonon lifetime has been realized using Matthiessen rule as \cite{kaviany_2014} 
\begin{equation}
    \frac{1}{\tau_{\lambda}} = \frac{1}{\tau_{\lambda}^N} + \frac{1}{\tau_{\lambda}^U} + \frac{1}{\tau_{\lambda}^I}
\end{equation}
where $\tau_{\lambda}^N$, $\tau_{\lambda}^U$ and $\tau_{\lambda}^I$ are phonon lifetimes corresponding to the normal, umklapp and isotope scattering respectively.

Generally, in harmonic approximation, phonon lifetimes are infinite whereas, anharmonicity in a crystal gives rise to a phonon self energy  $\Delta\omega_{\lambda}$ + $i\Gamma_{\lambda}$. The phonon lifetime has been computed from imaginary part of the phonon self energy as $\tau_{\lambda}$ = $\frac{1}{2\Gamma_{\lambda}(\omega_{\lambda})}$ from\cite{Togo}
\begin{widetext}
\begin{equation}
    \Gamma_{\lambda}(\omega_{\lambda}) = \frac{18\pi}{\hbar^2}\sum_{\lambda'\lambda''}\Delta\left( \textbf{q}+\textbf{q}'+\textbf{q}'' \right)\mid \Phi_{-\lambda\lambda'\lambda''}\mid ^{2}\{(n_{\lambda'}+n_{\lambda''}+1)\delta(\omega-\omega_{\lambda'}-\omega_{\lambda''}) + (n_{\lambda'}-n_{\lambda''})[\delta(\omega+\omega_{\lambda'}-\omega_{\lambda''}) - \delta(\omega-\omega_{\lambda'}+\omega_{\lambda''})]\}
\end{equation}
\end{widetext}
where $n_\lambda$ = $\frac{1}{exp(\hbar\omega_{\lambda}/k_{B}T)-1}$ is the phonon occupation number at the equilibrium. $\Delta\left(\textbf{q}+\textbf{q}'+\textbf{q}'' \right)$ = 1 if $\textbf{q}+\textbf{q}'+\textbf{q}'' = \textbf{G}$, or 0 otherwise. Here \textbf{G} represents reciprocal lattice vector. Integration over \textbf{q}-point triplets for the calculation is made separately for normal (\textbf{G} = 0) and umklapp processes (\textbf{G} $\neq$ 0). For both direct method and RTA, scattering of phonon modes by randomly distributed isotopes \cite{Togo} are also incorporated for comparison. The isotope scattering rate, using second-order perturbation theory, is given by Shin-ichiro Tamura \cite{Tamura} as
\begin{equation}
\resizebox{1.0\hsize}{!}{$
    \frac{1}{\tau_{\lambda}^{I}(\omega)} = \frac{\pi \omega_{\lambda}^{2}}{2N}\sum_{\lambda'} \delta\left(\omega - \omega'_{\lambda} \right) \sum_{k} g_{k}|\sum_{\alpha}\textbf{W}_{\alpha}\left(k,\lambda \right)\textbf{W}_{\alpha}^{*}\left(k,\lambda \right)| ^{2} 
$}
\end{equation}
where $g_k$ is mass variance parameter, defined as 
\begin{equation}
    g_{k} = \sum_{i} f_{i} \left( 1 - \frac{m_{ik}}{\overline{m}_k}\right)^{2}
\end{equation}
$f_i$ is the mole fraction, $m_{ik}$ is relative atomic mass of $i$th isotope, $\overline{m}_k$ is the average mass = $\sum_{i} f_{i} m_{ik}$ and $\textbf{W}$ is polarization vector. The database of the natural abundance data for elements \cite{Laeter} is used for the mass variance parameters.

For consistency check, we first simulate the lattice thermal conductivity ($\kappa_L$) of crystalline rhombohedral GeTe at 300 K using RTA method and compare the value of $\kappa_L$ with the results of the work done by D. Campi et al.\cite{Campi2017}. $\kappa_L$, obtained from our study using RTA, is 2.29 W/mK using the PBE functional in the DFT framework, which is in good agreement with the results of D. Campi et al.\cite{Campi2017} with $\kappa_L$ = 2.34 W/mK. After this consistency check, we use the DFT and Boltzmann transport equation (BTE) to get the lattice thermal conductivity of GeTe at the temperature regime studied in this work (322 K $\leqslant$ $T$ $\leqslant$ 503 K). 

Figure \ref{fig:kappa_expt_theory} and Table \ref{tab:table_kappa_compare} present the various contributions of thermal conductivity, obtained theoretically, superimposed with the experimental data. Starting from first principles calculations, both direct solution and RTA method are used to solve the LBTE to get the lattice thermal conductivity. Since $\kappa_L$ is anisotropic along hexagonal c axis and a-b axes, the average lattice thermal conductivity is calculated as $\kappa_{av}$ = $\frac{2}{3}$$\kappa_x$ + $\frac{1}{3}$$\kappa_z$ \cite{Mizokami, Campi2017}. Figure \ref{fig:kappa_expt_theory} and Table \ref{tab:table_kappa_compare} show that the resulting theoretical $\kappa_{tot}$ for the direct solution of LBTE is slightly overestimated compared to the experimental results, specifically at the higher temperatures ($T$ $>$ 322 K). However, the direct method is found to capture the trend of $\kappa(T)$ quite well. Consistently, slightly lower values of $\kappa$ are found due to the incorporation of the effect of phonon mode scattering due to isotopes. On the other hand, the results of thermal conductivity, obtained by solving Boltzmann transport equation under the relaxation time approximation (RTA), are found to be slightly underestimated as compared to the experimental results. However, as T $>$ 372 K, RTA results seem to be in better agreement with the experimental data and the difference between experimental and RTA goes down from $\approx$ 19 $\%$ at 322 K to $\approx$ 6 $\%$ at 503 K. The trend of $\kappa(T)$ obtained through RTA, alike the direct solution, is found to be in good agreement with the experimental trend. This trend of $\kappa(T)$ implies that umklapp scattering seems to dominate the phonon-phonon scattering at higher temperatures, as predicted by experiments. 

Previously, the hole concentration of GeTe is found to be  6.24 $\times$ 10$^{19}$ cm$^{-3}$, indicating a significant role of vacancies for the reported overestimation of the $\kappa(T)$, obtained via direct solution of LBTE. Indeed, D. Campi et al. \cite{Campi2017} found a considerable amount of lowering of $\kappa_L$ of crystalline GeTe at 300 K due to the vacancy present in the sample.
\onecolumngrid
\begin{widetext}
\vspace*{-0.5cm}
\begin{figure}[H]
    \centering
    \includegraphics[width=0.48\textwidth, angle=-90]{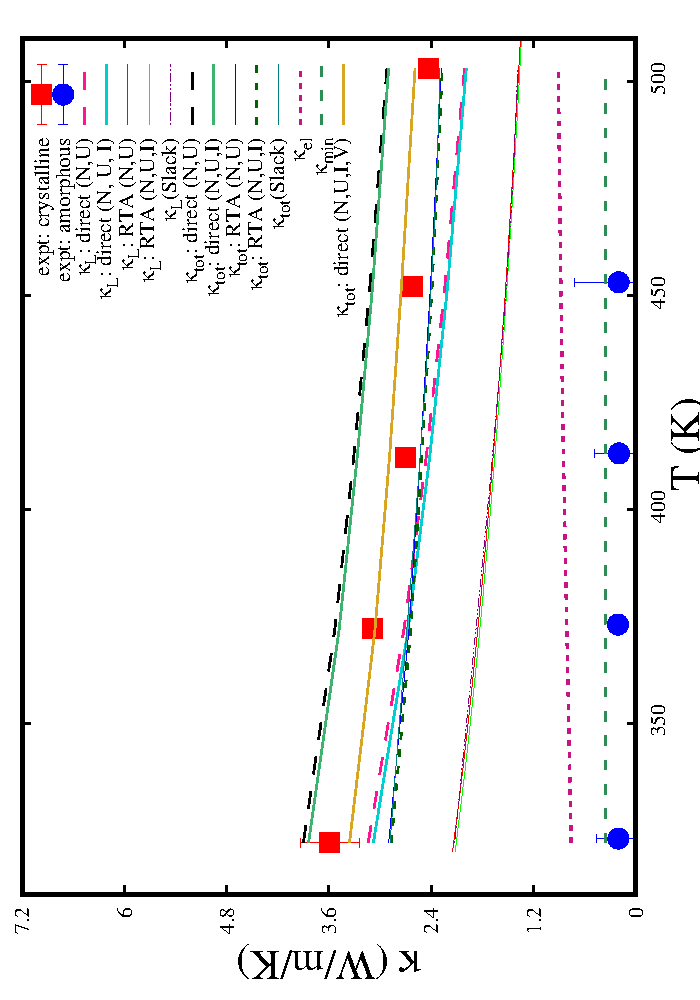}
    \caption{Experimental and theoretical thermal conductivity ($\kappa$) of GeTe as a function of temperature are presented in a fast change process. Phase change behavior is realized in the studied temperature range: 322 K $\leqslant$ T $\leqslant$ 503 K. Electronic ($\kappa_{el}$) and phonon contributions ($\kappa_L$) to the total thermal conductivity ($\kappa_{tot}$) are shown. $\kappa_L$ is evaluated using RTA, direct solution of LBTE and Slack model. N, U, I and V define normal, umklapp, isotope and vacancy scattering processes respectively.}
    \label{fig:kappa_expt_theory}
\end{figure}
\end{widetext}

\onecolumngrid
\begin{widetext}
\vspace*{-0.8cm}
\begin{table}[H]%The best place to locate the table environment is directly after its first reference in text
\caption{\label{tab:table_kappa_compare}%
Experimental and theoretical total thermal conductivity ($\kappa$) of the crystalline R3m-GeTe is presented as a function of temperature (T). The lattice contribution of thermal conductivity ($\kappa_L$) is taken as an average of $\kappa_x$ and $\kappa_z$ as $\kappa_{av}$ = (2$\kappa_x$ + $\kappa_z$)/3 (see text). The unit of $\kappa$ is Wm$^{-1}$K$^{-1}$. N, U, I and V represent normal, umklapp, isotope and vacancy scattering processes respectively.}
\begin{ruledtabular}
\begin{tabular}{cccccccc}
T (K) & Expt. & Direct (N,U) & Direct (N,U,I) & Slack & RTA(N,U) & RTA(N,U,I) & Direct (N, U, I, V)\\
\colrule
322 & 3.59 & 3.90 & 3.84 & 2.88 & 2.90 & 2.87 & 3.36\\
372 & 3.09 & 3.53 & 3.48 & 2.69 & 2.66 & 2.64 & 3.06\\
412	& 2.70 & 3.31 & 3.27 & 2.53 & 2.53 & 2.51 & 2.89\\
452 & 2.62 & 3.12 & 3.09 & 2.40 & 2.41 & 2.39 & 2.75\\
503 & 2.43 & 2.92 & 2.90 & 2.30 & 2.28 & 2.27 & 2.59\\
\end{tabular}
\end{ruledtabular}
\end{table}
\end{widetext}
The phonon scattering rate by vacancy defects are prescribed by C. A. Ratsifaritana et al. \cite{ratsifaritana} as
\begin{equation} {\label{eqvacancy}}
    \frac{1}{\tau_{V}} = x \left(\frac{\Delta M}{M}\right)^{2}\frac{\pi}{2}\frac{\omega^{2}g(\omega)}{G'}
\end{equation}
where, $x$ is the density of vacancies, $G'$ denotes the number of atoms in the crystal and $g(\omega)$ is the phonon density of states (PDOS). Using vacancies as isotope impurity, 
C. A. Ratsifaritana et al \cite{ratsifaritana} denoted mass change $\Delta M$ = 3$M$, where $M$ is the mass of the removed atom. Eq. \ref{eqvacancy} states that the phonon-vacancy relaxation time is temperature independent. D. Campi et al. \cite{Campi2017} used this phonon-vacancy scattering contribution for a GeTe sample with hole concentration of 8 $\times$ 10$^{19}$ cm$^{-3}$ and found an almost $\approx$ 15.6 $\%$ reduction of $\kappa_L$ at 300 K. As the hole concentration is almost similar to that of our work (6.24 $\times$ 10$^{19}$ cm$^{-3}$) , in conjunction with the fact that $\tau_{V}^{-1}$ is temperature independent, we estimate an overall $\approx$ 15.6 $\%$ decrement of $\kappa_{L}(T)$ throughout the temperature range studied as an effect of phonon-vacancy scattering. We find that the estimated $\kappa$, incorporating the vacancy contribution, in addition to the normal, umklapp and isotope scattering, agrees excellently with the experimental data for the whole temperature range in our study. This exercise strongly depicts the significant participation of the scattering between phonons and vacancy defects at high temperatures.

It is well known that the RTA, although describes the depopulation of phonon states well, fails to rigorously account the repopulation of phonon states \cite{Giorgia}. While at low temperatures, the applicability of RTA can be questioned due to the dominance of momentum conserving normal scattering and almost absence of umklapp scattering of phonons, in the high temperature regime of our study, RTA is found to be a good trade off between accuracy and the computational cost to describe the experimental results. This is primarily because of the higher number of scattering events at higher temperature which ensures an isothermal repopulation of the phonon modes \cite{Giorgia2}. However, the difference from the total solution of LBTE exists due to the over resistive nature of the scattering rates that effectively lowers the value compare to the direct solution. This feature has been discussed in literature  \cite{Giorgia2, Bosoni}. The reason of this underestimation is that the RTA treats both umklapp and normal scattering processes as resistive while the momentum conserving normal scattering processes do not equally contribute to the thermal resistance as that of the umklapp scattering \cite{Bosoni}.  

Simple phenomenological models can also serve a fast and efficient way to decipher the underlying physical mechanism. The Slack model \cite{Morelli2006} expresses the lattice thermal conductivity, when $T>\Theta_{D}$ and the heat conduction happens mostly by acoustic phonons, starting from the analytical expression of the relaxation time related to umklapp processes as:
\begin{equation}
\kappa_{L}=C\frac{\overline{M}\,\Theta_{D}^{3}\,\delta}{G^{2}\,n_{c}^{2/3}\,T}
\end{equation}
with:
\begin{equation}
C=\frac{2.43\times10^{-6}}{1-0.514/G+0.228/G^{2}}
\end{equation}
where $n_{c}$ is the number of atoms per unit cell, $\delta^{3}$ is volume per atom ($\delta$ is in angstrom in the relation), $\overline{M}$ is average atomic mass of the alloy and $G$ is the Gruneisen parameter (see Table \ref{table:parameters_cahill_slack}). This relation between lattice thermal conductivity and Gruneisen parameter in a solid is valid within a temperature range where only interactions among the phonons, particularly, anharmonic umklapp processes are dominant \cite{Morelli2006}. E. Bosoni et al. \cite{Bosoni} found a good agreement between the lattice thermal conductivity of crystalline GeTe coming from the full solution of BTE and that of the Slack model at room temperature.  Lattice thermal conductivity due to Slack model ($\kappa_L$(Slack)) is presented in Fig \ref{fig:kappa_expt_theory}. Total thermal conductivity is then realized by adding $\kappa_L$(Slack) with $\kappa_{el}$. We retrieve an almost identical trend of both lattice ($\kappa_L$) and total thermal conductivity ($\kappa_{tot}$) in Slack model as that of the RTA based solutions. This almost identical values of $\kappa_L$(Slack) and $\kappa_L$(RTA) depicts that the optical phonons contribute very little to $\kappa_L$(RTA) as $\kappa_L$(Slack) takes into account only acoustic phonon contributions. The values obtained for $\kappa_{tot}$(Slack) are found to be lower than the experimental data  for $T$ $<$ 412 K and gradually seem to agree well for $T$ $\geqslant$ 412 K  (Fig \ref{fig:kappa_expt_theory} and  Table \ref{tab:table_kappa_compare}).

Though RTA based solutions underestimate the experimental data, the trend of $\kappa(T)$, which is almost identical to phenomenological Slack model, is what needs attention to elucidate the underlying heat transport mechanism at high temperatures. Further, RTA based solutions are straightforward and can reveal the distinct role of each contributing parameters appearing in Eq.\ref{equation_kl}. Consequently, in the following sections, we systematically study the thermal transport properties of GeTe using both frequency and temperature variations using the results obtained from RTA solutions.

\subsection{Variation of lattice thermal conductivity ($\kappa_L$) with phonon frequency: Contribution of acoustic and optical modes}
 
To investigate lattice thermal conductivity ($\kappa_L$) in a more comprehensive manner, we calculate the cumulative lattice thermal conductivity as a function of phonon frequency for the high temperature regime defined as \cite{Togo,Mizokami}
\begin{equation}
    \kappa_L^c = \int_0^{\omega} \kappa_{L} (\omega')d\omega'
\end{equation}
\begin{figure}[H]
    \centering
    \includegraphics[width=0.5\textwidth]{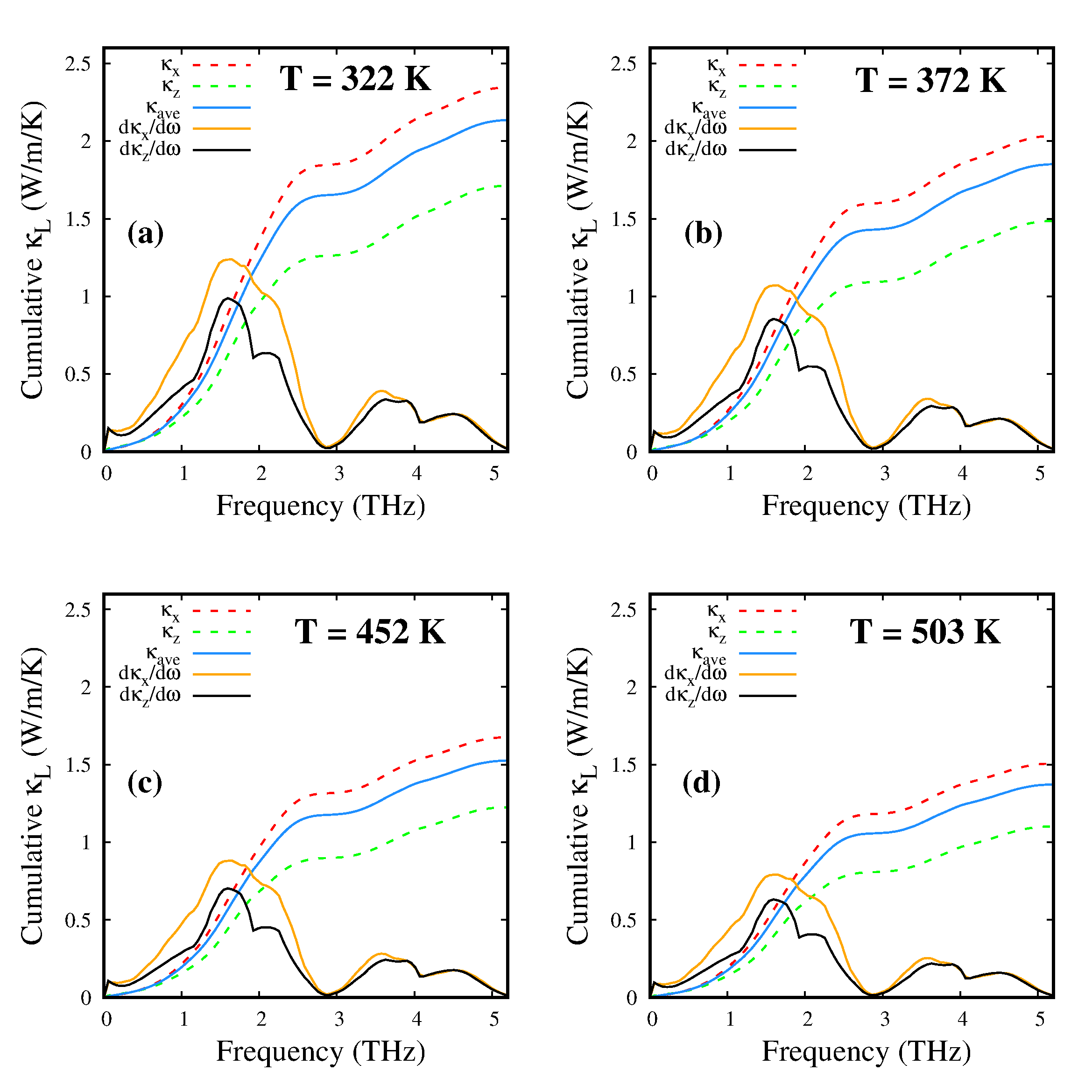}
    \caption{Cumulative lattice thermal conductivities ($\kappa_L$) of crystalline GeTe are presented as a function of frequencies at four different temperatures: (a) T = 322 K, (b) T = 372 K, (c) T = 452 K and (d) T = 503 K. Cumulative $\kappa_L$, computed within the RTA framework, along hexagonal c axis ($\kappa_z$), along its perpendicular direction ($\kappa_x$) and their average $\kappa_{av}$ = (2$\kappa_x$ + $\kappa_z$)/3 are shown. The derivatives of $\kappa_z$ and $\kappa_x$ with respect to frequencies are also shown for each temperature.}
    \label{fig:kappa_cumulative_theory}
\end{figure}
where $\kappa_{L}$ ($\omega'$) is defined as \cite{Togo,Mizokami}
\begin{equation}
    \kappa_{L}(\omega') \equiv \frac{1}{NV_0} \sum_{\lambda} C_{\lambda} \textbf{v}_{\lambda} \otimes \textbf{v}_{\lambda}\tau_{\lambda} \delta (\omega'-\omega_{\lambda}) 
\end{equation}
with $\frac{1}{N}$ $\sum_{\lambda} \delta(\omega' - \omega_{\lambda})$ is weighted density of states (DOS).

Figure \ref{fig:kappa_cumulative_theory} shows the cumulative $\kappa_L$ of crystalline rhombohedral GeTe, along hexagonal c axis ($\kappa_z$), along its perpendicular direction ($\kappa_x$) and their average $\kappa_{av}$ = (2$\kappa_x$ + $\kappa_z$)/3, as a function of phonon frequencies for four different temperatures using RTA framework. The derivatives of the cumulative values of $\kappa_z$ and $\kappa_x$ with respect to frequencies are also shown for each temperature. It is found that the lattice thermal conductivity ($\kappa_L$) of GeTe is anisotropic with the value along $z$, parallel to the c axis in the hexagonal notation, is found to be smaller with respect to that of the $xy$ plane for the whole range of temperature studied. This picture is consistent with the recent theoretical findings \cite{Campi2017}. More details to the anisotropic aspect of $\kappa_L$ will be discussed in the next subsection. 

Figure \ref{fig:kappa_cumulative_theory} shows some distinct features in the cumulative lattice thermal conductivity ($\kappa^c_L$) of GeTe as a function of both phonon frequency and temperature. As the temperature is increased, $\kappa^c_L$ is found to saturate at gradually 
\begin{table}[H]%The best place to locate the table environment is directly after its first reference in text
\caption{\label{tab:table_cumulative_kappa_contrib}%
Relative contributions of acoustic and optical modes to the total lattice thermal conductivity ($\kappa_L$) for different temperatures. The unit of $\kappa_L$ is W/mK.}
\begin{ruledtabular}
\begin{tabular}{cccc}
T (K) & $\kappa_L$ (RTA) & Contribution of & Contribution of \\
& & acoustic modes ($\%$) & optical modes ($\%$)\\
\colrule
322 & 2.14 & 77.1 & 22.9 \\
372 & 1.85 & 77.3 & 22.7 \\
452 & 1.53 & 77.1 & 22.9 \\
503 & 1.37 & 77.4 & 22.6 \\
\end{tabular}
\end{ruledtabular}
\end{table}
lower values, indicating gradual decrement of the lattice thermal conductivity with temperature. Further, the derivatives of $\kappa^c_L$ with respect to phonon frequencies indicate the density of heat carrying phonons with respect to the phonon frequencies \cite{Linnera} and their contribution to the $\kappa^c_L$. We note that this density of modes go to zero at a frequency where $\kappa^c_L$ reaches a plateau. This frequency ($\sim$ 2.87 THz) has been found to be identical to the frequency where phonon DOS marks the separation between acoustic and optical modes ($\sim$ 96 cm$^{-1}$ = 2.88 THz in Fig \ref{PDOS}). This correspondence imply that the phonon density of states play a crucial role as a deciding factor to the $\kappa_L$. While in Fig \ref{fig:kappa_cumulative_theory}, a significantly higher contribution of these modes correspond to $\kappa_x$ is observed compared to that of the $\kappa_z$ in the acoustic frequency regime (frequency $<$ 2.87 THz), an almost equal contribution of d$\kappa_x$/d$\omega$ and d$\kappa_z$/d$\omega$ are found in the optical frequency regime (frequency $>$ 2.87 THz).

To get a clear quantitative picture, we evaluate the separate contributions of acoustic and optical modes to the $\kappa_L$. The values of $\kappa^c_L$ for phonon frequencies $<$ 2.87 THz have unambiguously been considered to be the contribution from acoustic modes while the same for frequencies $>$ 2.87 THz have been taken as optical modes contribution. Table \ref{tab:table_cumulative_kappa_contrib} shows the percentage contributions of both acoustic and optical modes to the lattice thermal conductivity ($\kappa_L$) as a function of temperature. We observe that a dominant 77 $\%$ of the contribution comes from acoustic modes compared to only around 23 $\%$ from the optical modes for the whole temperature range studied. 

\subsection{Anisotropy of lattice thermal conductivity ($\kappa_L$) of crystalline $GeTe$}
\begin{figure}[H]
    \centering
    \includegraphics[width=0.5\textwidth]{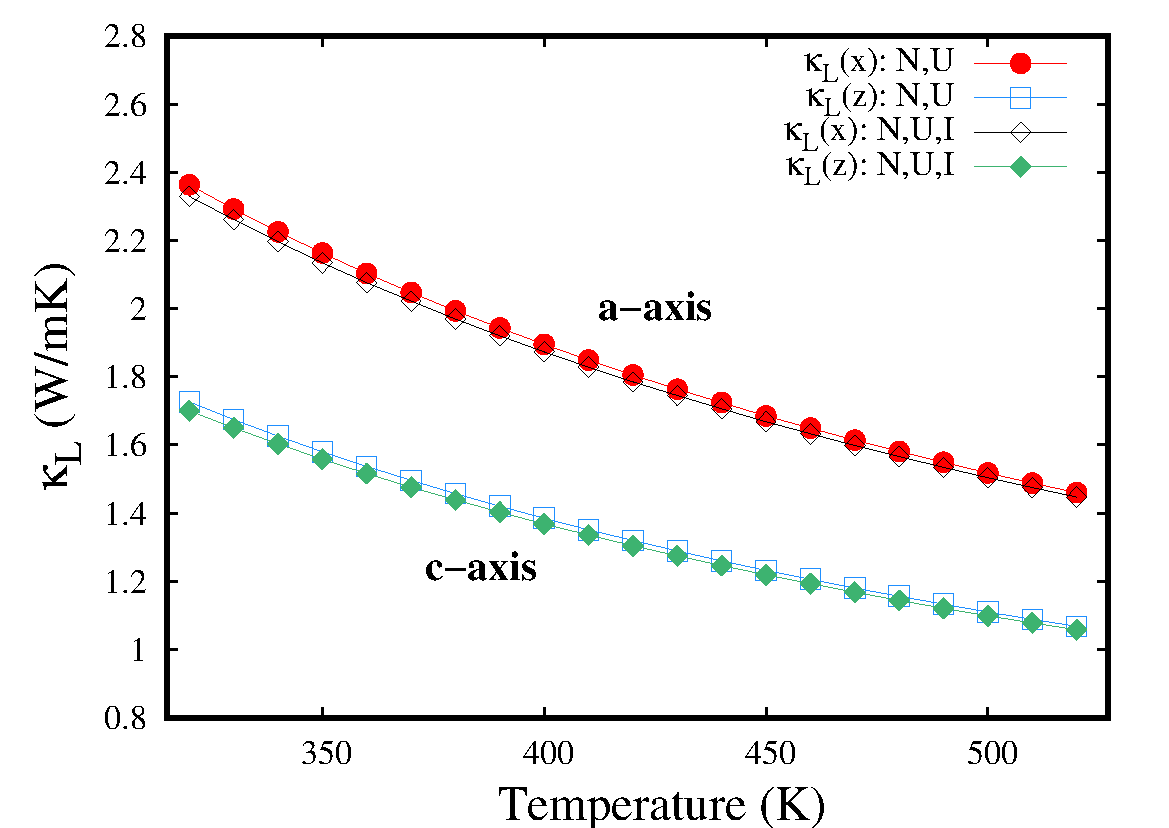}
    \caption{Lattice thermal conductivity ($\kappa_L$) of GeTe along $z$ direction ($\kappa_L(z)$, along the c axis in the hexagonal notation) and along $x-y$ direction (($\kappa_L(x)$, along the a axis) are presented as a function of temperature. N, U, I denote normal, umklapp and isotope scattering effects respectively.}
    \label{fig:kappa_x-z}
\end{figure}

As mentioned in the previous subsection, the lattice thermal conductivity ($\kappa_L$) of crystalline GeTe shows anisotropic behavior. The resulting $\kappa_L$ along $z$ direction ($\kappa_L(z)$), parallel to the c axis in the hexagonal notation (along axis c in Fig \ref{fig:1}) is found to be smaller than that of the $xy$ plane (a-b plane in Fig \ref{fig:1}),($\kappa_L(x)$), as shown in Fig \ref{fig:kappa_x-z}.

\begin{figure}[H]
    \centering
    \includegraphics[width=0.5\textwidth]{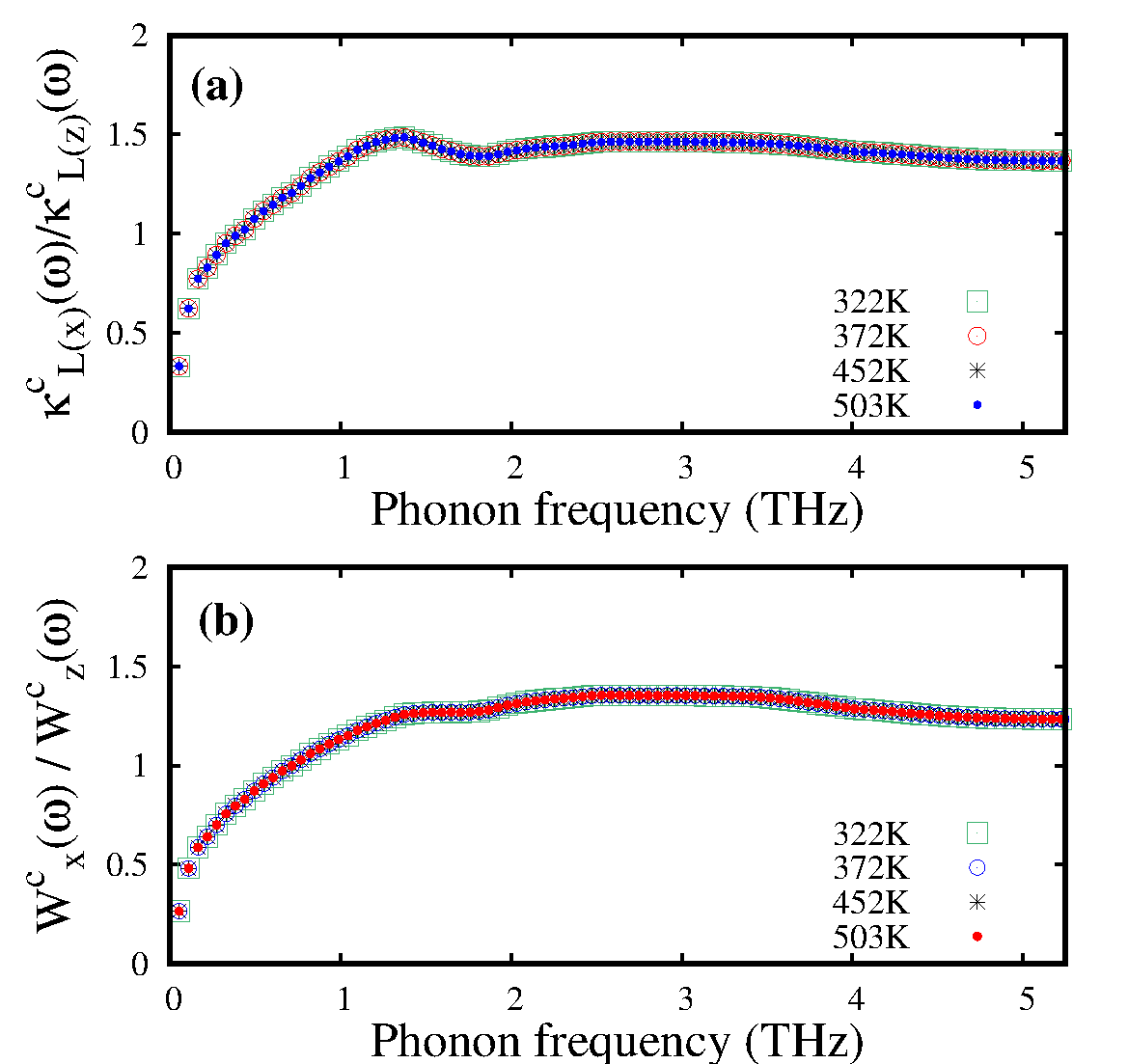}
    \caption{(a) Ratios of $x$ and $z$ components of cumulative lattice thermal conductivities, $\kappa^{c}_{L(x)}(\omega)$/$\kappa^{c}_{L(z)}(\omega)$ of crystalline GeTe as a function of phonon frequencies are presented for four different temperatures. (b) Ratios of $x$ and $z$ components of cumulative direct vector products of group velocities, $W^{c}_x$($\omega$)/$W^{c}_z$($\omega$) of crystalline GeTe as a function of phonon frequencies for four different temperatures.}
    \label{fig:kappa_anisotropy}
\end{figure}

To investigate the anisotropy, we study the ratio $\kappa^{c}_{L(x)}(\omega)$/$\kappa^{c}_{L(z)}(\omega)$, where $\kappa^{c}_{L(x)}(\omega)$ and $\kappa^{c}_{L(z)}(\omega)$ represent the cumulative lattice thermal conductivities along 
$xy$ plane (a-b plane in Fig \ref{fig:1}) and $z$ direction (along c axis in Fig \ref{fig:1}) respectively. Figure \ref{fig:kappa_anisotropy}.(a) shows this ratio as a function of phonon frequencies for the four different temperatures. We observe that despite the wide temperature range studied, the shape of the curves remain independent of the temperature. The ratio $\kappa^{c}_{L(x)}(\omega)$/$\kappa^{c}_{L(z)}(\omega)$, initially starts with a low value, becomes maximum to 1.5 at around 1.4 THz. With further increase of frequencies, the ratio decreases and settles at a value of 1.37.

The temperature independence of the anisotropy ratio $\kappa^{c}_{L(x)}(\omega)$/$\kappa^{c}_{L(z)}(\omega)$ can be further understood by studying the cumulative outer product of the phonon group velocities $\textbf{v}_\lambda \otimes \textbf{v}_\lambda$, defined as

\begin{equation}
    W^{c}(\omega) \equiv \frac{1}{NV_0} \sum_{\lambda} \textbf{v}_\lambda \otimes \textbf{v}_\lambda \delta(\omega - \omega_{\lambda})
\end{equation}
and $W^{c}_x$($\omega$)/$W^{c}_z$($\omega$), which is the ratio between the cumulative outer product of the phonon group velocities along $x$ and $z$ direction.  
Since phonon group velocities are almost temperature independent, following that feature, we find that the ratio of $W^{c}_x$($\omega$)/$W^{c}_z$($\omega$) is strongly correlated with $\kappa^{c}_{L(x)}(\omega)$/$\kappa^{c}_{L(z)}(\omega)$. From zero frequency, $W^{c}_x$($\omega$)/$W^{c}_z$($\omega$) starts increasing and reaches the ratio 1 around 0.7 THz, close to that of the $\kappa^{c}_{L(x)}(\omega)$/$\kappa^{c}_{L(z)}(\omega)$ (0.4 THz). Further increasing frequency increases the anisotropy between the in-plane ($xy$) and out-of the plane ($z$) components of the cumulative outer product of phonon group velocities and finally saturates to a value of 1.23 at higher frequencies which is also comparable to the saturation value of $\kappa^{c}_{L(x)}(\omega)$/$\kappa^{c}_{L(z)}(\omega)$, that is 1.37. Thus, the anisotropy associated with the phonon group velocities, or more specifically, the cumulative outer product of phonon group velocities determines the anisotropy in the cumulative lattice thermal conductivity.

\subsection{Variation of lattice thermal conductivity ($\kappa_L$) with temperature: Phonon lifetime and phonon mean free path}

Recalling Equation \ref{equation_kl}, in an alternate way to understand the frequency dependence of the different parameters that contribute to the lattice thermal conductivity ($\kappa_L$), we study the variation of modal heat capacity ($C_\lambda$), phonon group velocity ($\textbf{v}_\lambda$) and the phonon lifetime ($\tau$) as a function of phonon frequency for the high temperature regime studied in this work. The dominant contribution of phonon group velocities ($\textbf{v}_\lambda$) are found to arise from the acoustic phonons and the optical modes are found to exhibit substantially lower group velocities than the former. Furthermore, as expected, increasing temperatures does not change the dependence of group velocities on phonon frequencies. Similar behavior is expected from the modal heat capacity (C$_\lambda$) as a function of phonon frequencies for different temperatures. Indeed, C$_\lambda$ stays nearly constant with a small $\sim$ 2-5$\%$ deviation (2$\%$ for 503 K and 5$\%$ for 322 K) from $k_B$ which is consistent with the classical limit of C$_\lambda$ at high temperatures.  

\begin{figure}[H]
    \centering
    \includegraphics[width=0.5\textwidth]{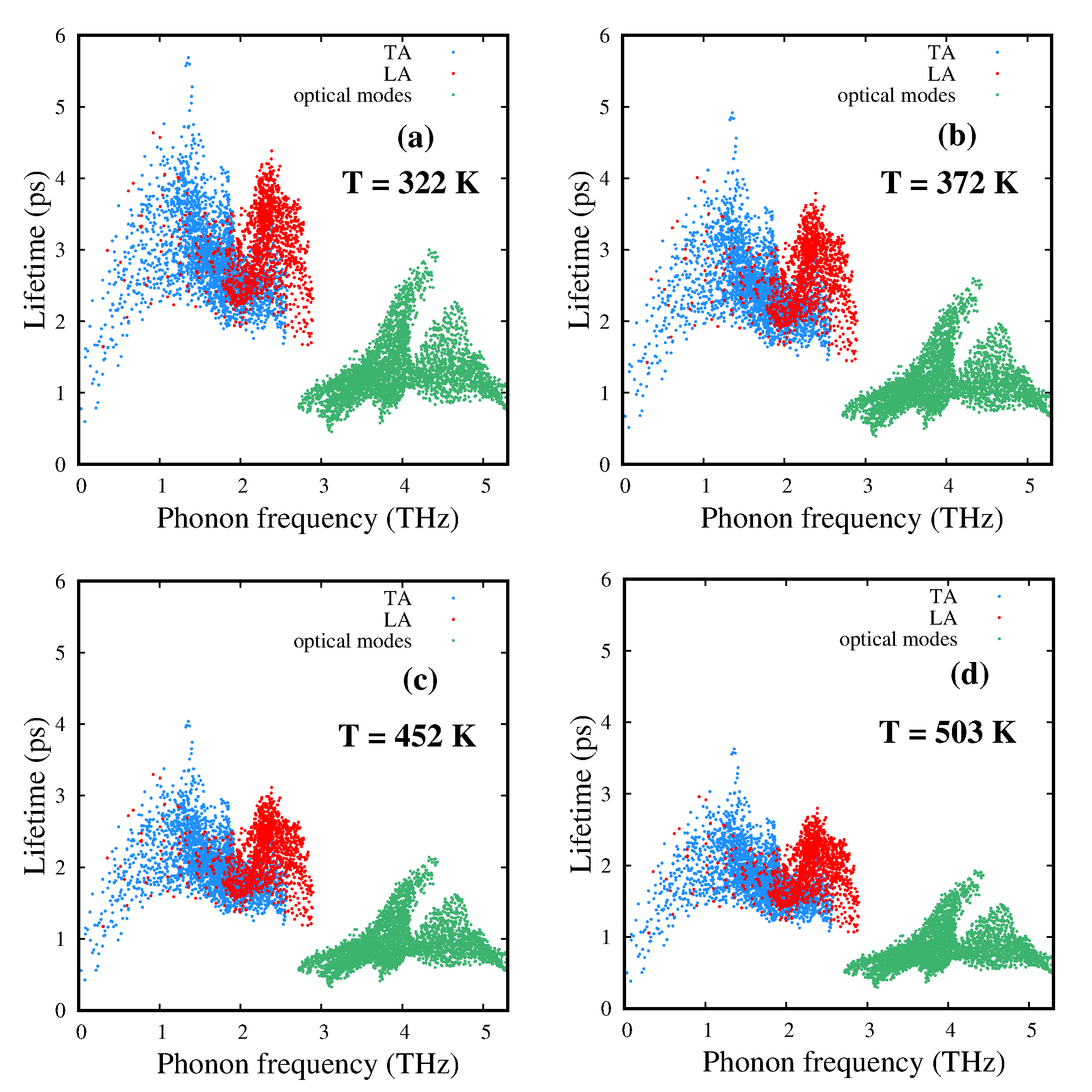}
    \caption{Phonon lifetimes of crystalline GeTe are shown as a function of phonon frequencies for four different temperatures: (a) T = 322 K, (b) T = 372 K, (c) T = 452 K and (d) T = 503 K. Phonon lifetimes due to transverse acoustic (TA), Longitudinal acoustic (LA) and optical modes are presented in blue, red and green points respectively.}
    \label{fig:kappa_lifetime_theory}
\end{figure}

This almost non varying patterns of phonon group velocities and mode heat capacity with temperature prompt us to look closely on the frequency variation of phonon lifetimes ($\tau_\lambda$) at these temperatures. We note here that $\tau_\lambda$ defines modal phonon relaxation time or phonon lifetime with $\lambda$ denotes each mode. Figure \ref{fig:kappa_lifetime_theory} depicts the phonon lifetimes, coming from TA, LA and optical modes as a function of phonon frequency. We observe that increasing phonon frequency from 0 THz gives an initial rise and then a quick decay of lifetimes within $\sim$ 3 THz and with relatively smaller values afterwards. This clearly indicates that the dominant contribution of phonon lifetime and in turn, the lattice thermal conductivity is coming from the acoustic modes, which operate in the frequencies $<$ 2.87 THz.

\begin{figure}[H]
    \centering
    \includegraphics[width=0.5\textwidth]{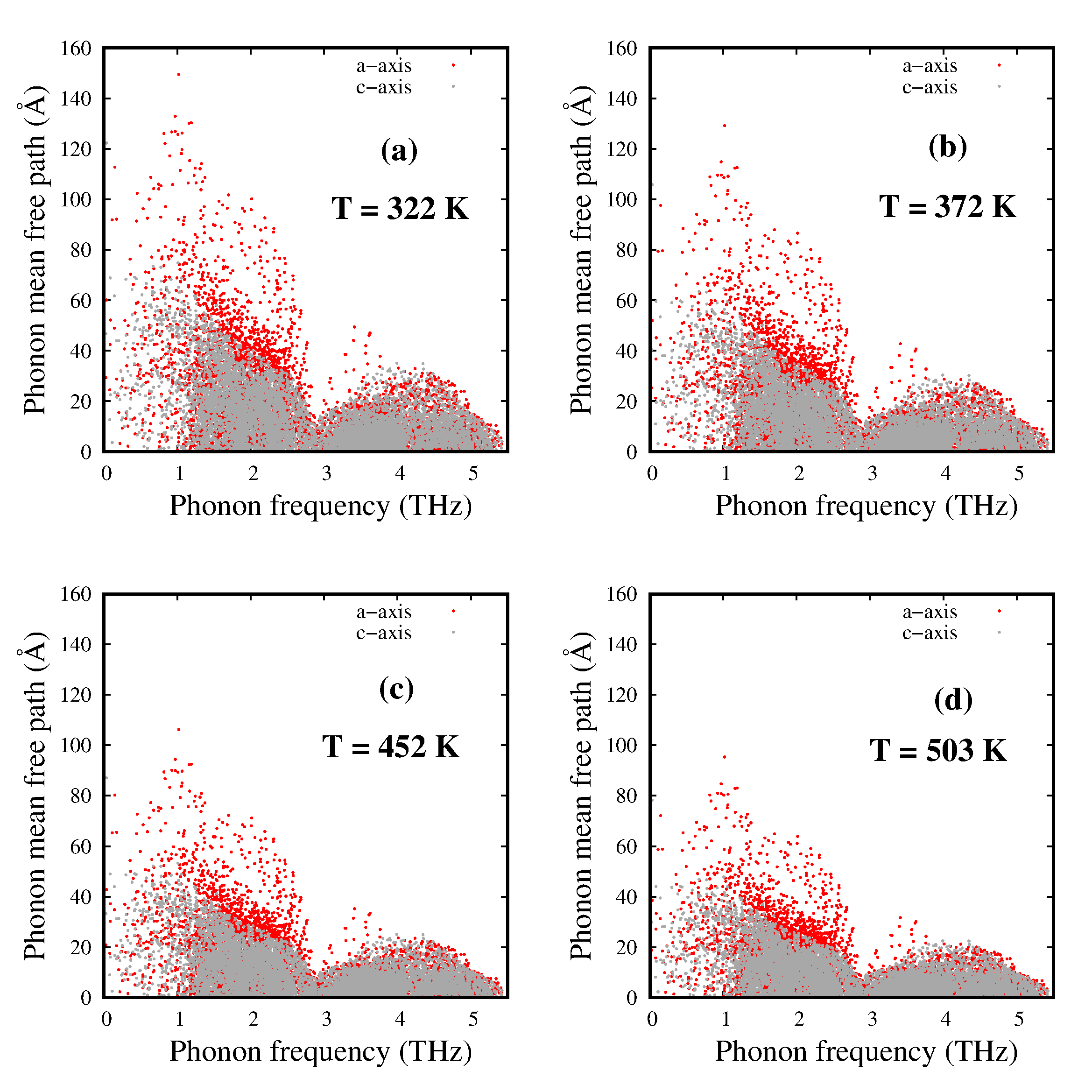}
    \caption{Phonon mean free paths of crystalline GeTe are shown as a function of phonon frequencies for four different temperatures: (a) T = 322 K, (b) T = 372 K, (c) T = 452 K and (d) T = 503 K. Phonon mean free paths along a-axis and c-axis are represented in red and gray dots respectively.}
    \label{fig:mfp_a_c}
\end{figure}
 Considering the temperature variation, we find that acoustic modes induce smaller values of phonon lifetimes with increasing temperature (Fig \ref{fig:kappa_lifetime_theory}). Two prominent observations evolve through this: (a) the trends in phonon lifetimes ($\tau_\lambda$) as a function of frequency reassure the fact that acoustic phonons are the dominant carriers of heat which contributes to $\kappa_L$, (b) as the phonon lifetime is directly proportional to $\kappa_L$, the reduced contribution of phonon lifetime with increasing temperature directs towards a gradual decrement of $\kappa_L$ with temperature. The gradually reducing values of phonon lifetime with increasing temperature can be understood more prominently considering the mean free path picture. The modal mean free path of phonons ($\Lambda_\lambda$) can be written as 
\begin{equation}
    \Lambda_\lambda = \textbf{v}_\lambda \tau_\lambda
\end{equation}
The transport of heat through phonons in the diffusive regime ($\Lambda$ $\ll$ $L$, $L$ = linear dimension of the medium of travelling phonons) undergoes several scattering processes namely scattering by electrons, other phonons, impurities or grain boundaries \cite{kaviany_2014}. At high temperatures, phonon-phonon scattering dominates along with impurity scattering to some extent. It is well understood that anharmonic coupling on thermal resistivity leads to $\Lambda$ $\propto$ $1/T$ at high temperatures \cite{Kittel86}. To elaborate it further, we study the mean free paths of phonons ($\Lambda_\lambda$) as a function of phonon frequencies for different temperatures. As can be seen from Fig \ref{fig:mfp_a_c}, the mean free paths decay quickly within $\sim$ 3 THz with gradually increasing phonon frequency starting from 0 THz and saturate at lower values afterwards. This is precisely due to the dominance of the acoustic modes in thermal transport of GeTe. With increasing temperature, the mean free paths are observed to exhibit lower values validating $\Lambda$ $\propto$ $1/T$. The anisotropic nature of $\kappa_L$ has also been understood by means of the mean free paths along a and c axes of R3m-GeTe. Separate contributions  of the mean free paths along a-axis and c-axis are shown in Fig \ref{fig:mfp_a_c}. Throughout the temperature range studied, phonon mean free paths correspond to the a-axis show higher values compared to that of the c-axis, giving rise to an enhanced heat transfer along a-axis with higher values of $\kappa_L (x)$ compared to the c-axis with lower values of $\kappa_L (z)$. Thus, the anisotropic mean free path distribution of phonons is found to be the key reason for exhibiting an anisotropic heat transfer in crystalline R3m-GeTe.

\begin{figure}[H]
    \centering
    \includegraphics[width=0.5\textwidth]{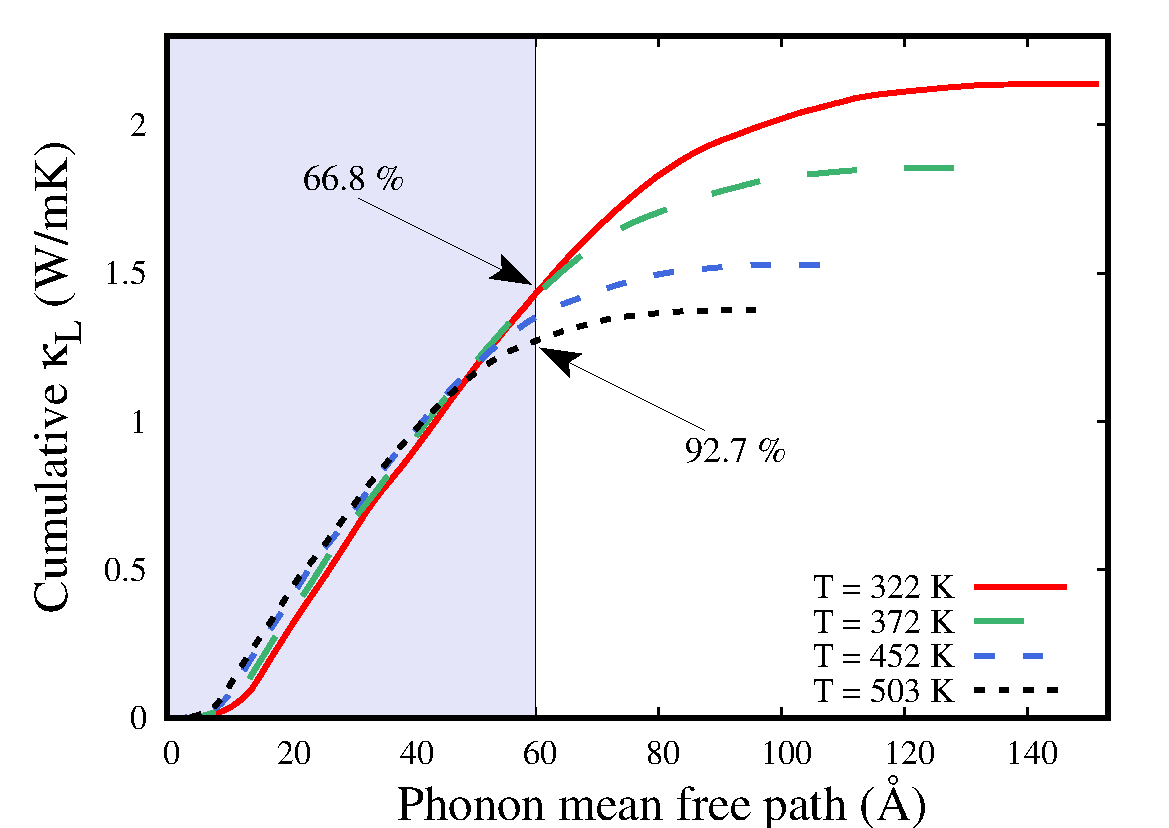}
    \caption{Variation of cumulative lattice thermal conductivity ($\kappa_L$) of crystalline GeTe with phonon mean free path is shown for four different temperatures: (a) T = 322 K, (b) T = 372 K, (c) T = 452 K and (d) T = 503 K. The contributions from the mean free paths $\leqslant$ 60 \AA{} to the $\kappa_L$ are mentioned for two extremes of temperatures.}
    \label{fig:kappa_mfp}
\end{figure}

To investigate the contribution of mean free paths of different lengthscales to $\kappa_L$, cumulative lattice thermal conductivity (considering the $\kappa_{ave}$ = (2$\kappa_x$ + $\kappa_z$)/3) is studied as a function of phonon mean free paths for the four different temperatures (Fig \ref{fig:kappa_mfp}). We observe from Fig \ref{fig:kappa_mfp} that the maximum values of the phonon mean free paths ($\Lambda_{max}$) gradually decrease with temperature (T = 322 K, $\Lambda_{max}$ $\sim$ 152 \AA; T = 372 K, $\Lambda_{max}$ $\sim$ 131 \AA; T = 452 K, $\Lambda_{max}$ $\sim$ 108 \AA; T = 503 K, $\Lambda_{max}$ $\sim$ 97 \AA), consistently with $\Lambda$ $\propto$ $1/T$. Moreover, a dominant contribution ($\approx$ 67 $\%$) to the lattice thermal conductivity is found to be coming from the phonon mean free paths $\leqslant$ 60 \AA. As temperature increases, the contributions from the phonon mean free paths $\leqslant$ 60 \AA{} are increased to almost 93 $\%$ (Fig \ref{fig:kappa_mfp}).

Increasing temperature, thus, manifests in a way of increasing phonon-phonon scattering processes which acts 
on the lowering of mean free path of phonons. With almost similar $\textbf{v}_\lambda$, the lowering of mean free path of phonons can be associated with the decrements of $\tau_\lambda$.

\subsection{Contribution of transverse and longitudinal acoustic phonons}
Up till now we understood the dominant contributions of the acoustic modes to the thermal transport mechanisms of crystalline rhombohedral GeTe. In this section, we further systematically discriminate the relative contributions of transverse and longitudinal acoustic phonons to the lattice thermal conductivity of crystalline GeTe for the temperature range investigated in this study. It is observed that the transverse modes (TA) contribute to almost 75 $\%$ while the longitudinal counterpart (LA) adds up the rest of 25 $\%$ to the lattice thermal conductivity for the whole temperature range.

\begin{figure}[H]
    \centering
    \includegraphics[width=0.35\textwidth,angle=-90]{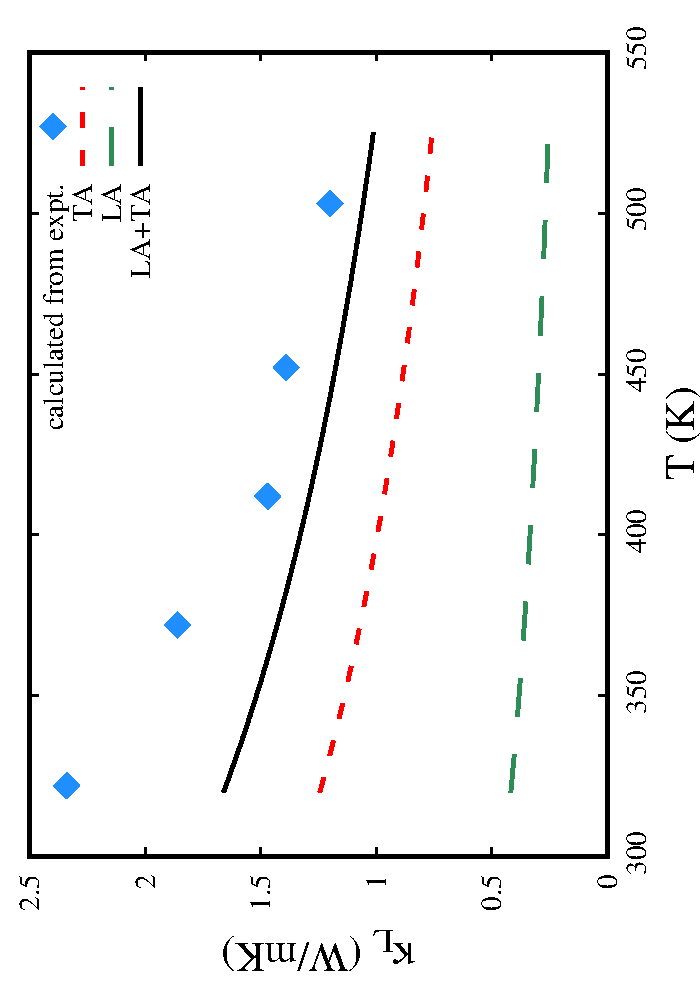}
    \caption{Lattice thermal conductivities, contributed solely from acoustic modes are shown as a function of temperature. `Experimental' values of acoustic $\kappa_L$ are obtained by subtracting $\kappa_{el}$ and optical mode contributed $\kappa_L$, from the experimental values of $\kappa$. RTA calculated $\kappa_L$ for transverse acoustic branch (TA), longitudinal acoustic branch (LA) and the total acoustic contribution (LA+TA) are also shown.}
    \label{fig:kappa_TA_LA_expt}
\end{figure}

Figure \ref{fig:kappa_TA_LA_expt} presents the lattice thermal conductivities, obtained via Ab-initio DFT calculations coupled with RTA, due to both transverse (TA) and longitudinal (LA) acoustic phonons as a function of temperature. For consistency, we also plot the $\kappa_L$ values for acoustic modes from the experimentally measured values of $\kappa$. This has been evaluated by subtracting the electronic thermal conductivity ($\kappa_{el}$), measured from the simulation, as well as the $\kappa_L$ due to optical phonons, computed from Ab-initio DFT and RTA, from the experimentally measured values of $\kappa$. Indicating a consistent picture from both experiment and theoretical calculations, as discussed earlier, an identical trend is found (Fig \ref{fig:kappa_TA_LA_expt}) between the data calculated from experiment and the simulated values of $\kappa_L$ due to acoustic phonons (TA+LA) despite the differences 
\

\begin{figure}[H]
    \centering
    \includegraphics[width=0.5\textwidth]{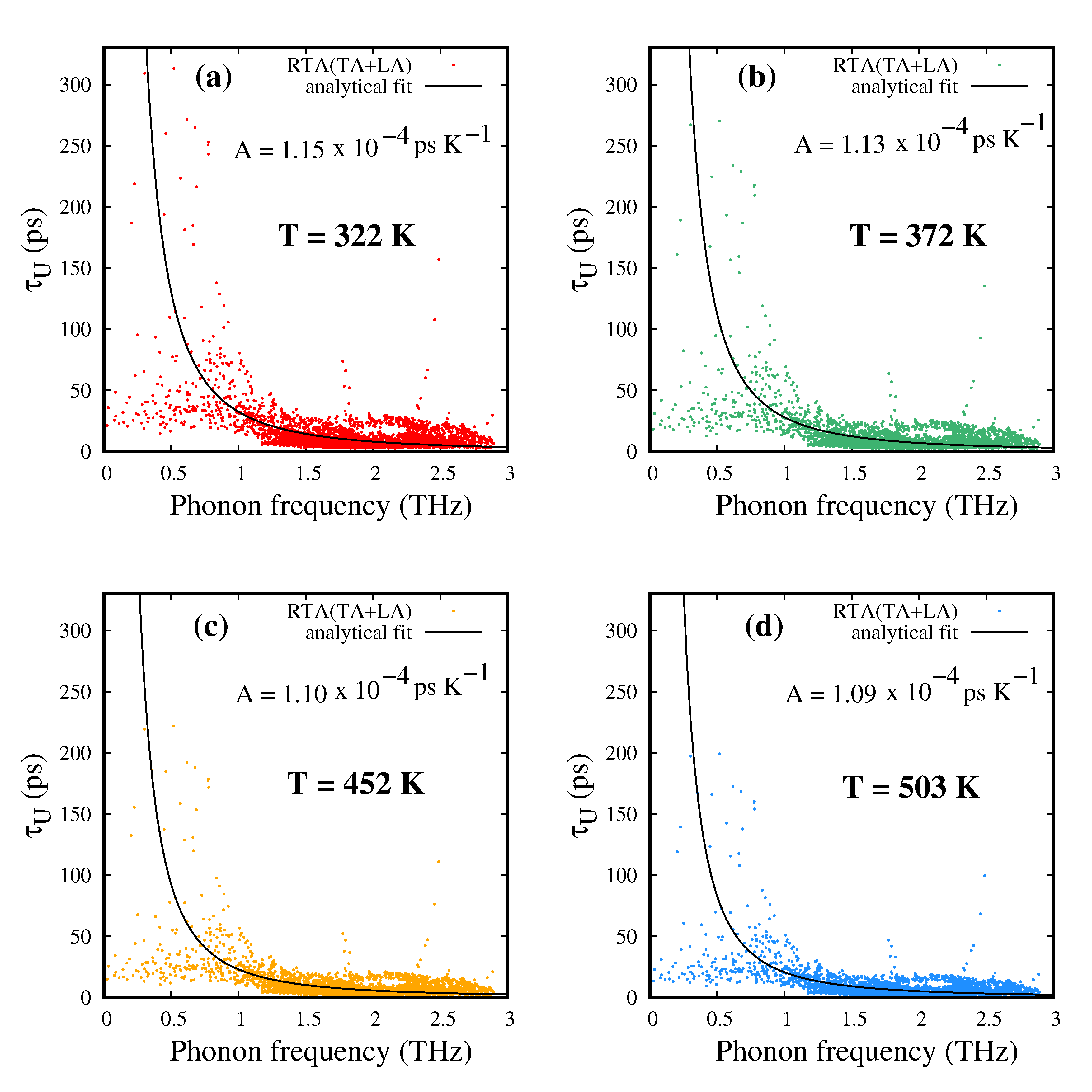}
    \caption{Phonon lifetimes of GeTe corresponding to umklapp scattering, along with the umklapp fitted parameter $A$ are shown as a function of phonon frequencies for four different temperatures: (a) T = 322 K, (b) T = 372 K, (c) T = 452 K and (d) T = 503 K.}
    \label{fig:umklapp_fit_lifetime_RTA}
\end{figure}
between the values, mostly for T $<$ 412 K.

Throughout this work, the trend of simulated $\kappa(T)$ using RTA has been found to be a straightforward and convenient framework to describe the temperature dependence of the experimental results of $\kappa$ for crystalline GeTe. Further, the exact similarity between RTA and the phenomenological Slack model, indicates that umklapp scattering plays an important and significant phonon-phonon scattering mechanism in the high temperature range ($T$ $\gg$ $\Theta_D$), $\Theta_D$ being the Debye temperature. To identify the umklapp scattering parameter involved in the process, we fit the phonon relaxation time correspond to the umklapp scattering, obtained from RTA, with the analytical expression given by Glen A. Slack et al. \cite{slack1964} 
%\onecolumngrid
%\begin{widetext}

%\end{widetext}

 \begin{equation}
    \tau_{U}^{-1} = AT\omega^{2}exp\left(-B\frac{\Theta_D}{T}\right) 
\end{equation} 
with $A$ $\propto$ $\frac{\hbar G^{2}}{\overline{M}c^{2}\Theta_{D}}$ and $B$ $\sim$ 1/3, where $\overline{M}$ is average atomic mass of the alloy and $G$ is the Gruneisen parameter. Being an empirical equation, it is necessary to find the parameter $A$ from the fitting procedure. Figure \ref{fig:umklapp_fit_lifetime_RTA} shows the phonon relaxation times as a function of phonon frequency along with the fitted curve for all the four temperatures. We observe that at higher frequencies the trend of phonon relaxation time indeed shows $\omega^{-2}$ dependence for all temperatures. The values of $A$, thus retrieved from the fitting, are found to be independent of temperature with an average value of 1.1$\times$ 10$^{-4}$ ps K$^{-1}$. 

This quantification, via the parameter $A$, serves as an important generic identification as this parameter can uniquely distinguish and compare the umklapp scattering processes for different crystalline materials ranging from $\Theta_D$ to an arbitrary high temperatures, within the operational regime of umklapp scattering process.

\section{Summary and Conclusions}{\label{section:summary}}

We have carried out a systematic experimental and theoretical study on the thermal conductivity variation of GeTe at high temperatures. The study involves fast and reversible phase change between amorphous and crystalline phases of GeTe. Modulated photothermal radiometry (MPTR) as well as the Lavenberg-Marquardt (LM) technique are employed to determine thermal conductivities of GeTe in both amorphous and crystalline phases as a function of temperature. Thermal boundary resistances, coming from both Pt-GeTe and GeTe-SiO$_2$ interfaces, have been accurately taken into account for measuring $\kappa$ experimentally. Van der Pauw technique as well as Boltzmann transport equations are solved for electrons to estimate electronic thermal conductivity within the constant relaxation time approximation (CRTA) framework. 

To compute lattice thermal conductivity ($\kappa_L$), first-principles density functional theory (DFT) is used and the solution to the linearized Boltzmann transport equation (LBTE) has been realized via both direct method and relaxation time approach (RTA). Normal, umklapp and isotope effects are included in computing the phonon relaxation time in these approaches. While the direct method is found to capture the trend of $\kappa(T)$ quite well, the values are a bit overestimated compared to the experimental data. The hole concentration of 6.24$\times$10$^{19}$ cm$^{-3}$, obtained using first principles calculations and BTE for electrons, necessitates the incorporation of phonon-vacancy scattering to estimate $\kappa_L$. Following a recent work \cite{Campi2017} on crystalline GeTe with almost same hole concentration, vacancy contribution is incorporated to estimate more realistic values of $\kappa$. Indeed, the estimate of $\kappa$ using direct method and adding the temperature independent phonon-vacancy scattering contribution, an excellent agreement is obtained between experimental and theoretical values of $\kappa$.

$\kappa$ computed from RTA, is also found to retrieve the trend of experimental $\kappa(T)$ quite well, especially at higher temperatures. However, the over-resistive nature of RTA due to the treatment of umklapp and normal scattering in equal footing, causes an underestimation of $\kappa$ compared to the experimental values. Nevertheless, the trend $\kappa(T)$ agrees well at higher temperatures. Cumulative lattice thermal conductivity is presented as a function of phonon frequencies for different temperatures. The density of heat carrying phonons or rather the phonon density of states plays a crucial role in determining $\kappa_L$.  Acoustic phonons emerge as the dominant ($\sim$ 77$\%$) contributor to the lattice thermal conductivity while transverse acoustic modes contribute almost 75$\%$ to the acoustic phonon transport. The anisotropy of lattice thermal conductivities of crystalline GeTe between in-plane and out of plane components are understood using the variation of cumulative outer product of phonon group velocities. Further, the anisotropy present in phonon mean free path along a-axis and c-axis are found to control the anisotropy in the heat transfer along these two axes. Phonon group velocities and mode heat capacities are observed to remain almost independent of temperature. Therefore, the temperature variation of $\kappa$ is attributed to the variation of phonon mean free path and consequently the variation of phonon lifetime with temperature. 

Phenomenological models are also found to be quite effective in describing and identifying the underlying physical mechanism of thermal transport. The experimental values of $\kappa$ in amorphous phase of GeTe are described using the minimal thermal conductivity model, proposed by Cahill et al. \cite{Cahill}. For crystalline phase data, Slack model \cite{Morelli2006} is also employed and it has been found to be strikingly similar with RTA based solutions. Both RTA based solutions as well as expression from phenomenological Slack model for GeTe indicate that the umklapp phonon phonon scattering is significant for the temperature regime studied in this work. Umklapp phonon relaxation time is found to obey $\omega^{-2}$ dependence at higher frequencies and therefore umklapp scattering parameter has been obtained, which remains almost constant for the whole temperature range studied. This complete experimental and theoretical exercise to elucidate the thermal conductivity of GeTe at high temperatures can further assist in improving the thermal management for other Te based phase change materials by understanding heat dissipation, localization and transport with more clarity.

\begin{acknowledgments}
This project has received funding from the European Union’s Horizon 2020 research and innovation program under Grant Agreement No. 824957 (“BeforeHand:” Boosting Performance of Phase Change Devices by Hetero- and Nanostructure Material Design).
\end{acknowledgments}

% The \nocite command causes all entries in a bibliography to be printed out
% whether or not they are actually referenced in the text. This is appropriate
% for the sample file to show the different styles of references, but authors
% most likely will not want to use it.
\nocite{*}

%\bibliography{ref}% Produces the bibliography via BibTeX.
%apsrev4-2.bst 2019-01-14 (MD) hand-edited version of apsrev4-1.bst
%Control: key (0)
%Control: author (8) initials jnrlst
%Control: editor formatted (1) identically to author
%Control: production of article title (0) allowed
%Control: page (0) single
%Control: year (1) truncated
%Control: production of eprint (0) enabled
%
\end{document}